\documentclass[apj]{emulateapj}
 \usepackage{rotating}
 \usepackage{amsmath} 
\usepackage{color}
\usepackage{hyperref}
\usepackage{tabularx}

\def\ltsima{$\; \buildrel < \over \sim \;$}
\def\simlt{\lower.5ex\hbox{\ltsima}}
\def\gtsima{$\; \buildrel > \over \sim \;$}
\def\simgt{\lower.5ex\hbox{\gtsima}}
\def\kpc{{\rm\,kpc}}

\def\pc{{\rm\,pc}}

\def\deg{^\circ}

\def\s{\ifmmode \widetilde \else \~\fi}
\def\={\overline}

\def\spose#1{\hbox to 0pt{#1\hss}}

\def\lta{\mathrel{\spose{\lower 3pt\hbox{$\mathchar"218$}}
     \raise 2.0pt\hbox{$\mathchar"13C$}}}
\def\gta{\mathrel{\spose{\lower 3pt\hbox{$\mathchar"218$}}
     \raise 2.0pt\hbox{$\mathchar"13E$}}}
\def\Dt{\spose{\raise 1.5ex\hbox{\hskip3pt$\mathchar"201$}}}    
\def\dt{\spose{\raise 1.0ex\hbox{\hskip2pt$\mathchar"201$}}}    

\def\dotsfill{\leaders\hbox to 1em{\hss.\hss}\hfill}

\def\Gyr{{\rm\,Gyr}}
\def\FeH{{\rm[Fe/H]}}

\DeclareMathOperator\erfc{erfc} 

\defcitealias{Martin2013}{M13} 

\shorttitle{Andromeda's dwarf galaxy detection limits}
\shortauthors{A. Doliva-Dolinsky et al.}

\begin{document}


\title{The PAndAS View of the Andromeda Satellite System. III. Dwarf galaxy detection limits}


\author{Amandine Doliva-Dolinsky$^{1}$, Nicolas F. Martin$^{1,2}$, Guillaume F. Thomas$^{3,4}$, Annette M. N. Ferguson$^{5}$, Rodrigo A. Ibata$^{1}$, Geraint F. Lewis$^{6}$, Dougal Mackey$^{7}$, Alan W. McConnachie$^{8}$, Zhen Yuan$^{1}$}
\email{amandine.doliva-dolinsky@astro.unistra.fr}

\altaffiltext{1}{Universit\'e de Strasbourg, CNRS, Observatoire astronomique de Strasbourg, UMR 7550, F-67000, France}
\altaffiltext{2}{Max-Planck-Institut f\"ur Astronomie, K\"onigstuhl 17, D-69117, Heidelberg, Germany}
\altaffiltext{3}{Instituto de Astrof\'isica de Canarias, Calle Vía L\'actea s/n, E-38206 La Laguna, Tenerife, Spain}
\altaffiltext{4}{Universidad de La Laguna, Avda. Astrof\'isico Fco. S\'anchez, E-38205 La Laguna, Tenerife, Spain}
\altaffiltext{5}{Institute for Astronomy, University of Edinburgh Royal Observatory, Blackford Hill, Edinburgh EH9 3HJ, UK}
\altaffiltext{6}{Sydney Institute for Astronomy, School of Physics, A28, The University of Sydney, NSW 2006, Australia}
\altaffiltext{7}{Research School of Astronomy and Astrophysics, Australian National University, Canberra, ACT 2611, Australia}
\altaffiltext{8}{NRC Herzberg Astronomy and Astrophysics, 5071 West Saanich Road, Victoria, BC, V9E 2E7, Canada}

\begin{abstract}
We determine the detection limits of the search for dwarf galaxies in the Pan-Andromeda Archaeological Survey (PAndAS) using the algorithm developed by the PAndAS team (\citealt{Martin2013}). The recovery fractions of artificial dwarf galaxies are, as expected, a strong function of physical size and luminosity and, to a lesser extent, distance. We show that these recovery fractions vary strongly with location in the surveyed area because of varying levels of contamination from both the Milky Way foreground stars and the stellar halo of Andromeda. We therefore provide recovery fractions that are a function of size, luminosity, and location within the survey on a scale of $\sim1\times1$\,deg$^2$ (or $\sim14\times14\kpc^2$). 
Overall, the effective surface brightness for a 50-percent detection rate range between 28 and 30 mag\,arcsec$^{-2}$. This is in line with expectations for a search that relies on photometric data that are as deep as the PAndAS survey. 
The derived detection limits are an essential ingredient on the path to constraining the global properties of Andromeda's system of satellite dwarf galaxies and, more broadly, to provide constraints on dwarf galaxy formation and evolution in a cosmological context.
\end{abstract}


\section{Introduction}

In cosmological models that include dark matter, hundreds of sub-halos of dark matter are orbiting around a central halo \citep{Klypin1999,Moore1999}. These sub-haloes are the place of birth of dwarf galaxies, but not all will "light up". The expected number of dwarf galaxies is highly sensitive to the cosmology \citep{Spergel2000,Bode2001} but also to diverse physical phenomena that can stop star formation, such as stellar feedback or the reionization \citep{Bullock2000,Somerville2002,Mashchenko2008,Wheeler2015}. Dark matter simulations with adding baryon physics are compared to dwarf galaxies observations in order to constrain those parameters \citep{Koposov2009,Kim2018,Nadler2019}.

The search for dwarf galaxies was revolutionized by large, homogeneous photometric surveys such as the Sloan Digital Sky Survey (SDSS; \citealt{SDSS2003}), the Panoramic System Telescope and Rapid Response System~1 $3\pi$ survey (Pan-STARRS1; \citealt{Pan-STARR12016}) or the Dark Energy Survey (DES; \citealt{DarkEnergySurvey12018}). Those new datasets are searched for dwarf galaxies thanks to detection algorithms that look for over densities of stars compatible with an old stellar population \citep{Belokurov2007,Martin2013,Laevens2015}. Thanks to these improvements, numerous new dwarf galaxies were detected. Indeed, before those surveys, there was a dozen known dwarf galaxies around the Milky Way (MW) with an absolute magnitude $M_{V} <-8.8\pm0.2$ (Draco; \citealt{Mateo1998}). By now, the number of known MW satellites is 59, with $M_{V} <-0.8\pm0.9$ (Virgo I, \citealt{Daisuke2016}). Even if the nature of those objects is sometimes debated \citep{Conn2018-1,Jerjen2018,Mutlu-Pakdil2018}, the increase in the number of known dwarf galaxies allows us to determine the dwarf galaxy detection limits of each survey, and so, to obtain more precise statistical estimations of the real number of dwarf galaxies around the MW. This was already done for SDSS, Pan-STARRS1 and DES searches \citep{Koposov2008,Drlica2020-1}, and used to derived an estimate of the luminosity function for the MW. 
	
The Local Group hosts another large galaxy that provides us with a different test sample for cosmological and galaxy formation models. Large photometric surveys also changed our knowledge on M31 and its satellites. Searches focused on the SDSS led to the discovery of 3 new dwarf galaxies around M31, as faint as $M_{V} =-8.1\pm0.5$ with And~X \citep{Zucker2004,Zucker2007,Bell2011}. The INT/WFC imaging of the surrounding of M31 led to the discovery of And~XVII \citep{Irwin2008}. But the main survey leading to the discovery of numerous new M31 dwarf galaxies is the Pan-Andromeda Archaeological Survey (PAndAS), a dedicated survey of the halo of the Andromeda galaxy. Searches within this survey led to the discovery of 19 satellite dwarf galaxies of M31 with $M_{V} <-5.9\pm0.7$ \citep{Richardson2011,McConnachie2008,Ibata2007,Martin2006}. For this survey, an algorithm was created by the PAndAS team to search for stellar populations of dwarf galaxies in the full survey by looking for overdensities of stars both spatially and in the color-magnitude diagram \citep[CMD;][\citetalias{Martin2013} hereafter]{Martin2013}. Thanks to the algorithm, we can now determine the survey completeness limit. 

In this paper, we determine the detection limits of Andromeda's dwarf galaxies using the PAndAS survey. In parallel to similar studies already conducted on the Milky Way, this will provide a second independent test for cosmological or galaxy formation simulations, using a different satellite system. Recovery fractions of dwarf galaxies are obtained by adding artificial galaxies to the PAndAS photometric catalogue and by determining whether they are detected by blindly running the search algorithm developed by the PAndAS team \citepalias{Martin2013}.

This paper is structured as follows: Section \ref{Data}, briefly describes PAndAS and its data. Section \ref{Methods}, details the generation of artificial dwarf galaxies and how we determine their recovery fractions. Section \ref{Results} presents the modeled dwarf galaxy completeness for each field of the PAndAS survey. The impact of the galaxy's distance and metallicity is also investigated. Finally, we discuss our results in Section~\ref{Discussion}.

\section{Preliminaries} \label{Data}

The Pan-Andromeda Archaeological Survey (PAndAS) was a Large Program conducted from 2008 to 2011 at the Canada-France-Hawai Telescope (CFHT), using the wide-field imager MegaCam to map the surroundings of the Andromeda galaxy \citep{McConnachie2009,McConnachie2018}. Combined with previous observations obtained through PI time, it includes over 400 fields of $\sim$1 deg$^2$ each and probes a region within $\sim150\kpc$ of M31 and $\sim50\kpc$ of its companion galaxy M33. Each field was observed in both the $g$ and $i$ band for photometry that includes the brightest three magnitudes of the red giant branch (RGB) at the distance of M31. The survey is presented in detail in \cite{McConnachie2018}, along with the data reduction and catalogue creation steps, and we refer the reader to this publication for the full description of the catalogues that we use here. In a nutshell, these reach median, $5\sigma$ g and i depths of 26.0 and 24.8, respectively \citep{Ibata2014}.


The algorithm to generate dwarf galaxies, described and used below, requires two field-specific ingredients: a model of the photometric uncertainties and a model of the completeness. \citet{Thomas2021} presents models for these two components which are determined carefully for a small set of fields and propagated to the full survey by scaling these initial models with an anchor magnitude that corresponds to the magnitude at which the uncertainties are equal to 0.1\,magnitude.

\begin{figure}
\begin{center}
\includegraphics[width=\hsize]{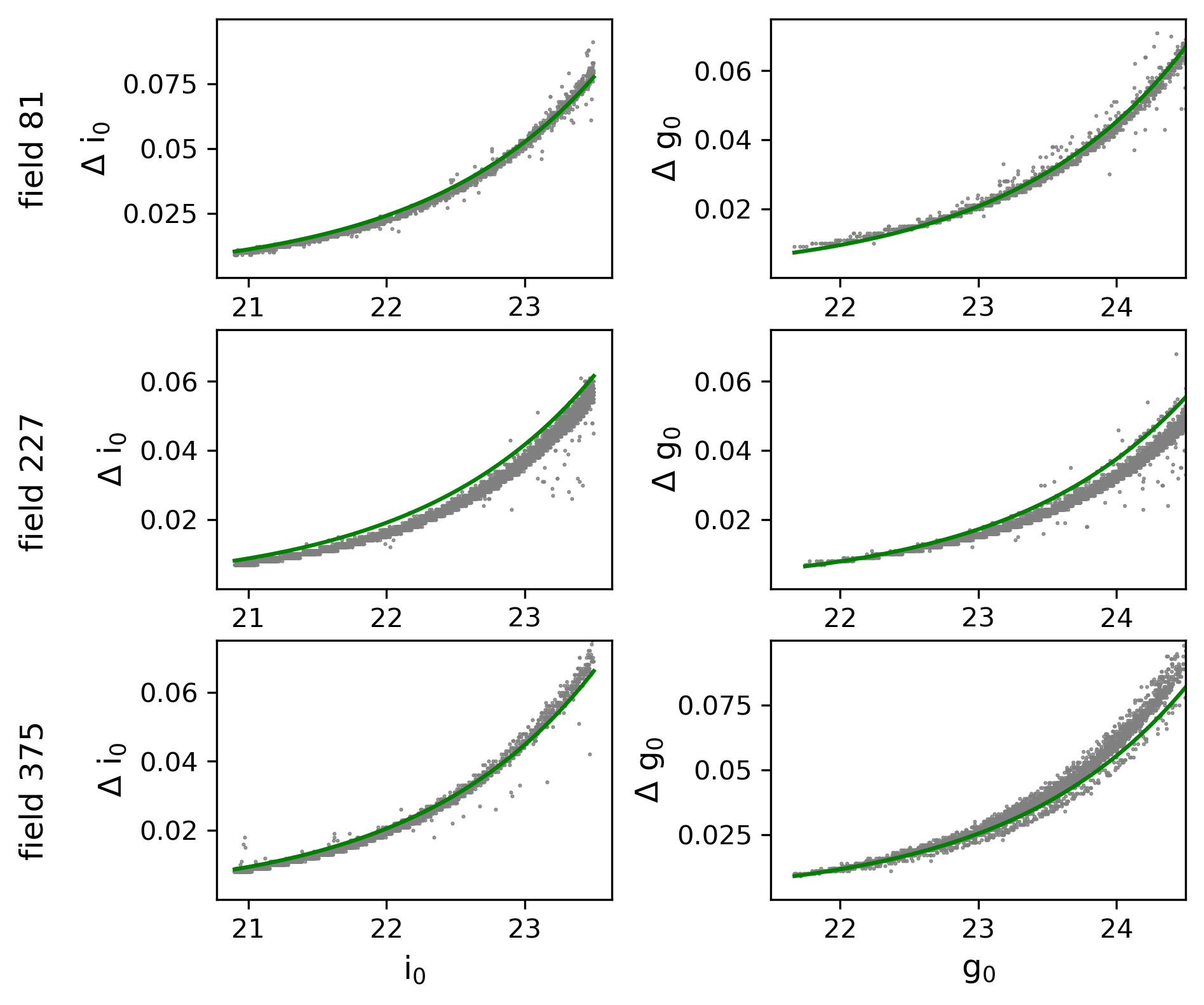}
\caption{\label{uncertainty}  Uncertainties in the $i$ (left) and $g$ (right) magnitudes for 3 typical fields of the survey. The exponential fits (green lines), reasonably follow the photometric uncertainty distributions of each field.}
\end{center}
\end{figure}

For the model of the photometric uncertainties, we model them with an exponential model as a function of magnitude (see \citealt{Ibata2007} for more detail). The photometric uncertainties in the $i$ and $g$ bands are shown in Figure~\ref{uncertainty} for 3 different fields of the survey with the corresponding models. The exponential fits describe the photometric uncertainties reasonably well, despite small differences that arise between fields.

For the completeness, we use the work of \citet{Thomas2021} who determined, independently, the $i$- and $g$-band completeness of the PAndAS data by comparison with regions of deep photometry observed with the Hubble Space Telescope (HST) data. This process is based on 14 HST fields that are $\sim$2.5-3 magnitudes deeper than PAndAS and propagated to the full survey via changes in the anchoring magnitude, as described above.

\section{Methods}
\label{Methods}
\subsection{Generating artificial dwarf galaxies}
\label{Generatingdwarfgalaxy}

To determine the dwarf-galaxy detection limits of the PAndAS survey we need to generate artificial dwarf galaxies that can be added to the PAndAS photometric catalogue to measure their probability of being detected. For any artificial dwarf galaxy, we aim to generate a catalogue of fake positions and $g$ and $i$ magnitudes that, combined, create a system of known properties (total magnitude, metallicity, radius, distance, and/or age). 

The first step to generate an artificial dwarf galaxy is to determine the photometry of its individual stars. For simplicity, we assume that the artificial dwarf galaxy can be parametrized by a single age and metallicity. While this is clearly not accurate, the different stellar populations of the faint dwarf galaxies that will be simulated here are not clearly separated in the PAndAS photometry \citep{Martin2016}. For every artificial dwarf galaxy, we go through the following steps:

\begin{enumerate}
 \item For a given choice of metallicity, age, distance, and total magnitude in the $V$ band ($M_V$), we randomly draw stars from the corresponding luminosity function taken from the PARSEC library \citep{Marigo2017}, using the pre-2014 MegaCam photometric system and a Kroupa initial mass function \citep{Kroupa2001}. The corresponding isochrone from the same library provides the color that the randomly drawn stars must follow. Artificial stars are drawn with an absolute magnitude $-5 < M_i < 20$ until the total flux of the system reaches the target total magnitude $M_V$\footnote{MegaCam magnitudes are transformed into V-band magnitudes using the color equations presented in \cite{Ibata2014}.}. We then keep only bright stars as the selection box of the search algorithm will only use upper Red Giant Branch (RGB) stars with $i_0<23.5$. The exact shape of the horizontal branch is not modeled in detail as its stars are fainter than the magnitude limit used by the algorithm but their flux is properly taken into account to determine the total luminosity of the artificial dwarf galaxy.
 
  \item To best represent the PAndAS observations, we randomize the perfect magnitudes drawn from the isochrone and luminosity function by adding noise to the magnitudes, following the photometric uncertainties models of the field in which the artificial galaxy will be placed. The added observed photometric uncertainties broaden the locus of stars in the CMD. 
  
  \item Finally, we use the (in)completeness model determined by \citet{Thomas2021} to test for the observability of any star drawn in the artificial galaxy. Every star generated in the previous step is tested against the completeness model. For every star drawn from the isochrone and luminosity function, we draw random deviates between 0 and 1 and independently test those against the completeness at the star's g- and i-band magnitudes.
\end{enumerate}

\begin{figure*}
\begin{center}
\includegraphics[width=\hsize]{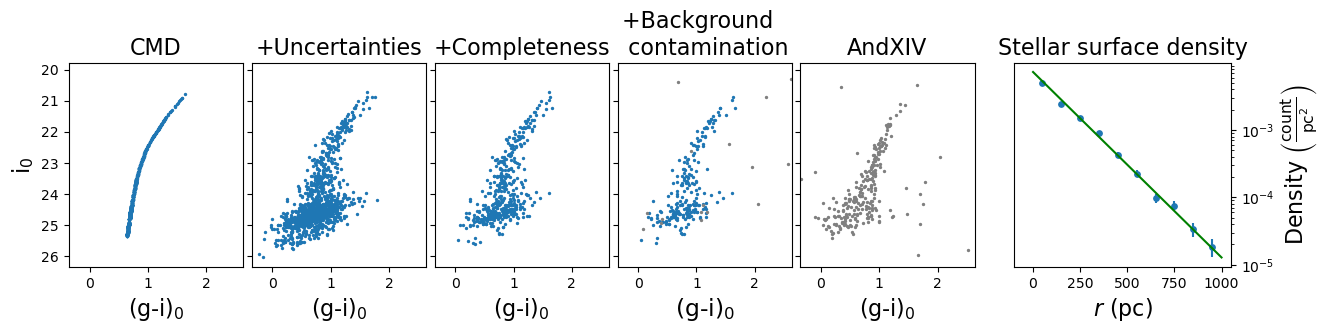}
\caption{\label{panneau_isochrone} CMD of the stars of an artificial dwarf galaxy with $M_V=-8.5$, $[\textrm{Fe}/\textrm{H}]=-1.7$, an age of 10 $\Gyr$, and $r_\mathrm{h}$=265 pc, as it is pushed through the different steps of the generation process. The first panel corresponds to the CMD of stars drawn from the chosen luminosity function and corresponding isochrone. As the search algorithm is only looking at RGB stars we only represent the bright stars that are of interest here. The second panel shows the CMD after taking the PAndAS photometric uncertainties into account. The CMD of the third panel folds in the incompleteness of the PAndAS data. Finally, the fourth panel shows the CMD of this artificial system within $2r_h$ overlaid on top of the typical field contamination from PAndAS (grey points). A direct comparison with the CMD within $2r_h$ of And~XIV (fifth panel) that has similar parameters shows the similarities between the two CMDs. The sixth panel displays the perfect agreement between the radial surface density of stars in the artificial dwarf galaxy (blue points) and the chosen radial density model (green line).}
\end{center}
\end{figure*}

The CMD of an artificial dwarf galaxy with $M_V=-8.5$, $\FeH=-1.7$, an age of 10 Gyr (and $r_\mathrm{h}=265 \pc$) is shown in Figure~\ref{panneau_isochrone} as it is pushed through the different steps of the process. We can see that the resulting CMD in the fourth panel, which is the one of an artificial dwarf galaxy added on a random location of the PandAS survey without any known satellite or stellar stream, is very similar to the CMD of the real dwarf galaxy And~XIV that has a similar total luminosity ($M_V=-8.5$; \citealt{Martin2016}) and is shown in the fifth panel. 

Now that the number of stars in the galaxy and their photometry is known, the next step is to determine their position with respect to the center of the galaxy. The radial stellar density is assumed to follow an exponential law with scale radius $r_\mathrm{e} =r_\mathrm{h}/1.68$, where $r_\mathrm{h}$ is the half-light radius of the system in pc. The probability density function of stars, $P(r)$, at radius $r$ given in pc, is 
\begin{equation}
    P\left(r\right)=\frac{1.68^2r}{2\pi r_\mathrm{h}^2}\exp\left(-1.68\frac{ r}{r_\mathrm{h}}\right).
\end{equation}

\noindent For every star in the artificial dwarf galaxy that was assigned a color and a magnitude in the previous step and passed the completeness test, we draw its radius from this probability distribution function and assign it a random angle between 0 and $2\pi$. These polar coordinates are then transformed into sky coordinates using the center of the artificial dwarf galaxy. The right-hand panel of Figure~\ref{panneau_isochrone} shows the perfect agreement between the radial density profile of stars in this test artificial dwarf galaxy and the chosen model density profile.

\subsection{Determining the recovery rate of artificial galaxies}
There are clear differences in the properties of observed stars throughout the PAndAS survey as, for instance, the contamination from MW foreground stars increases significantly towards the north \citepalias{Martin2013} or the density of M31 stellar halo stars changes with the density of stellar stream stars \citep{Ibata2014}. We therefore aim to determine the recovery fractions of artificial dwarf galaxies (and therefore the detection limits of dwarf galaxies in the survey) as a function of the location in the survey. A degree, at the distance of M31, corresponds to $\sim14\kpc$. This means that the $1\deg\times1\deg$ MegaCam field is a natural areal subdivision to consider since it corresponds to a sizable area of the M31 halo without being too large. The fact that there are subtle field-to-field differences in the photometry depth and completeness, despite the survey being very homogeneous, further comforts us in our choice to determine the artificial dwarf-galaxy recovery fractions on a field-to-field basis.

\subsubsection{Choice of parameters for the artificial dwarf galaxies}
We first need to carefully decide which parameters have the most significant impact on the detection limits so as to limit as much as possible the significant amount of computation required to determine the dwarf galaxy recovery fractions. 

We know that the parameters with the most impact on the detection of a dwarf galaxy (or absence therefore) are the size and the total luminosity of a system because these are directly related to the surface brightness of a dwarf galaxy \citep[e.g.][]{Koposov2009,Walsh2009,Drlica2020-1}. The distance to the system is very important for searches of MW dwarf galaxies but is not so important for distant systems. At the distance of M31, a change in distance of $\pm300\kpc$ only leads to a change to the location of stars in the CMD by $-1.1/+0.5$ magnitudes, respectively. 
While this is not subtle, the fact that PAndAS only observes red-giant branches that are sparsely populated, and so very noisy, means that a shift in distance is similar to a shift in metallicity and so the distance does not have a very significant impact on the recovery rate of a dwarf galaxy \citepalias{Martin2013}.

To limit the required calculations, we therefore consider the distance to an artificial system as a secondary parameter and, for the moment, assume all artificial dwarf galaxies are located at the distance of M31, with a distance modulus of 24.47 \citep{McConnachie2005}.

The range of interest for a dwarf galaxy's total magnitude is $-4.5 \lta M_V \lta -8.5$ \citep{Martin2016} and straddles the total magnitude of the faintest dwarf galaxy detections. In this regime, all known dwarf galaxies are invariably metal-poor with a metallicity contained between -1.5 and -2.3 \citep[][]{Tollerud2012,Collins2013}. In addition, the PAndAS photometry is not very sensitive to metallicity variations in this range as the tracks followed by RGB stars in this metallicity range almost overlap in the $(g-i,i)$ CMD. To save on unnecessary computing time, we therefore assume a fixed metallicity for all dwarf galaxies, with $\FeH=-1.7$.

Finally, we know that the age of the stellar populations in a dwarf galaxy have little impact on the color and magnitude of its RGB stars as long as they are at least moderately old ($\gta2\Gyr$). Consequently, we assume an age of $10\Gyr$ for all artificial dwarf galaxies \citep{Weisz2019}.

We check the impact of these three assumptions on the distance, the metallicity, and the age in \S~\ref{Results} below and confirm that it is indeed minimal, or can easily be modeled in the case of the distance.

In summary, for each field in the PAndAS footprint, we generate artificial galaxies with a metallicity $\FeH=-1.7$, an age of $10$Gyr, at a distance modulus of M31 ($m-M=24.47$). To test the impact of the total magnitude ($M_V$) and the size (parametrized by the half-light radius $r_h$ given in pc) of the system, we bin the $M_V$--$\log_{10}(r_\mathrm{h})$ plane over the generous range $-8.5<M_V<-4.5$ and $1.8<\log_{10}(r_\mathrm{h})<3$, with bin sizes of 0.25 and 0.1, respectively. From known M31 dwarf galaxies, we know that the transition from detected to undetected systems is within this range \citep{Martin2016}. For each of these bins, we generate an artificial galaxy with a random size and magnitude within the limits of this bin and ingest the photometry and position of its stars in the PAndAS catalogue at a random location within the considered field. Stars that fall in neighboring fields are kept but we remove any star of the artificial galaxy that falls in a CCD gap, in a hole between 2 fields or outside of the PAndAS footprint and we then test for the detection significance of this system (see below). This step is usually repeated 5 times so we can determine a recovery fraction for the studied magnitude-size bin and field. Ideally, we would like to ingest more than 5 artificial galaxies but, given the $\sim400$ fields and the number of bins in the magnitude-size plane, this already corresponds to about half a million ingested artificial galaxies and running the search algorithm 1000 times since ingesting all the artificial dwarf galaxies at once would lead to overlapping systems, change the properties of the survey, and bias the recovery rates. However, we will confirm below from a small subset of representative fields that ingesting only 5 artificial dwarf galaxies per $M_V$--$\log_{10}(r_\mathrm{h})$ bin and per field still yields reliable results.

\begin{figure*}
\begin{center}
\includegraphics[width=\hsize]{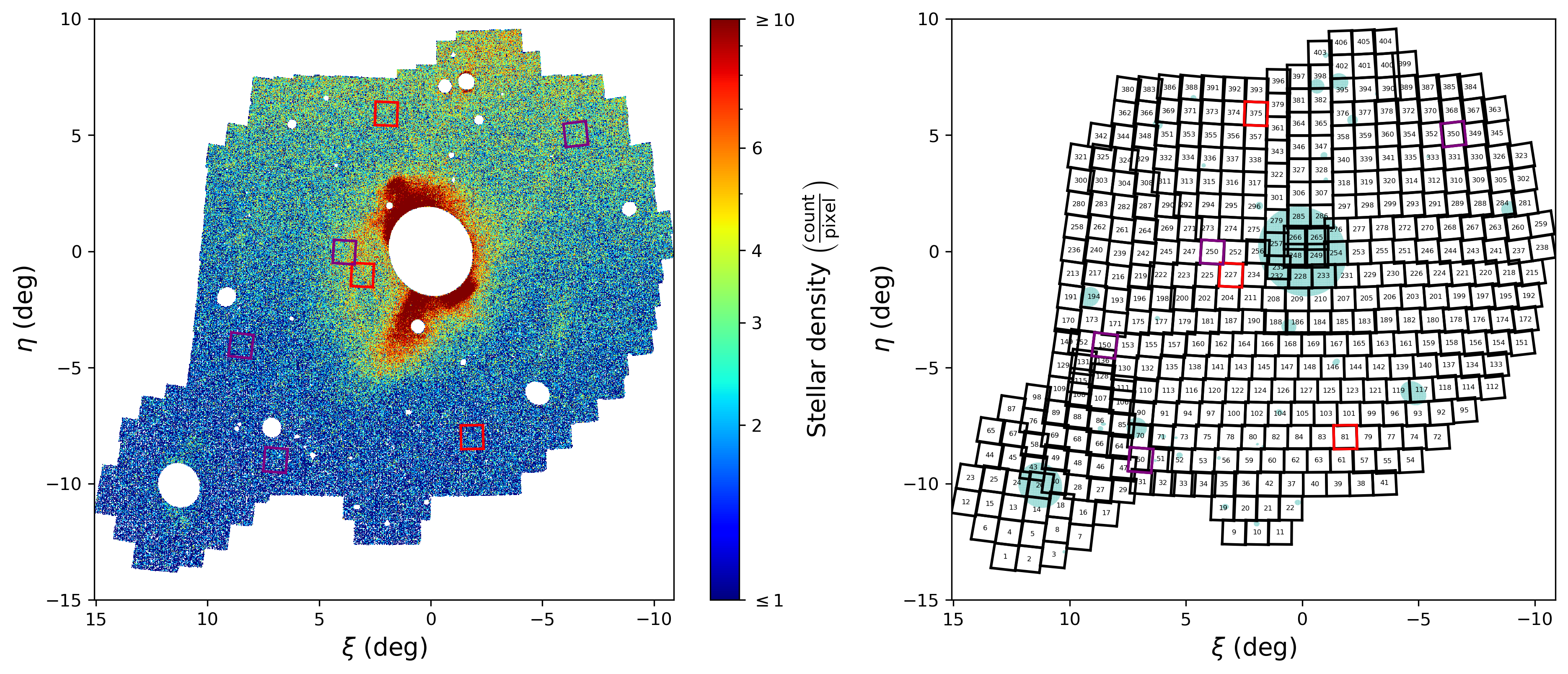}
\caption{\label{localdensity} Local density of PAndAS stars in the CMD selection box (left panel). The white ellipses correspond to the regions we mask because of known stellar systems and the three fields chosen to be representative of the various regions of the survey are highlighted in red. The additional 4 fields used to test the effect of the distance on the recovery fractions are represented in purple. The right-hand panel shows the distribution of PAndAS fields, following the numbering scheme of \citet{Ibata2014} and \citet{McConnachie2018}. Masked regions are here shown in cyan.}
\end{center}
\end{figure*}

Finally, we remove regions near known dwarf galaxies where the search algorithm would invariably return a detection, irrespective of the ingested system, and the regions close to M31 and M33, for which the search algorithm does not work well because of the complex mix of stellar populations \citepalias{Martin2013}. We therefore mask regions within $4r_h$ of all known dwarf galaxies, as well as within $\sim2$ and $\sim1$ degrees of M31 and M33, respectively. In the south of the survey, a group of background nearby elliptical galaxies with globular clusters that masquerade as M31 RGB stars also needs to be masked out. We choose not to mask M31's globular clusters as they are rarely recovered by the search algorithm and only one of these is detected in \citetalias{Martin2013} with a significance above 6. The resulting mask is shown in Figure~\ref{localdensity}. The detection limits of fields that are entirely within the masked regions are not determined and, for partially masked fields, we make sure to insert artificial dwarf galaxies in the non-masked part of the field.

\subsubsection{Set up of the search algorithm} 
To test for the recovery of an artificial galaxy, we use the algorithm developed by \citetalias{Martin2013} to search for dwarf galaxies in the PAndAS survey. The algorithm is computationally costly to run as it determines the likelihood of there being a dwarf galaxy at all locations of PAndAS on a $0.5\arcmin$ grid ($115\pc$ at the distance of M31), given the distribution of local RGB stars on the sky and in the color-magnitude space. The assumption is that this distribution can be well reproduced by three different components: one that simulates a metal-poor, compact dwarf galaxy, one that simulates the locally constant contamination of M31's halo stars, and a third part that represents the contamination from MW stars, following an empirically built model. We refer the reader to \citetalias{Martin2013} for the full description of the search algorithm and its parametrization but we mention here our choices for the model parameter grid that we let the algorithm explore. Ideally, it would be best to explore a broad range of choices for all parameters of the model but, as mentioned in \citetalias{Martin2013}, this can be quite costly and we restrict the exploration to likely model parameters.

\begin{table}
\begin{center}
\begin{tabular}{lccc}
  \hline
  Parameter & Minimal value & Maximal value & Step \\
  \hline
  $\FeH_\mathrm{halo}$ & $-1.3$ & $-0.6$  & 0.1 \\
    $\FeH_\mathrm{dw}$ & $-2.3$ & $-1.1$ & 0.3\\
    $r_\mathrm{h}$ & 0.5\arcmin & 3.5\arcmin & 1\arcmin\\
    $\log_{10}(N)$ & $-0.5$ & 4 & 0.5\\
  $\eta$ & 0 & 1 & 0.1 \\
  \end{tabular}
  \caption{Ranges of parameters used to run the search algorithm}
  \label{param}
  \end{center}
\end{table}

The five model parameters we focus on are:
\begin{itemize}
    \item the metallicity of the dwarf galaxy, $\FeH_\mathrm{dw}$;
    \item the half-light radius of the dwarf galaxy $r_\mathrm{h}$;
    \item the number of stars in the dwarf galaxy component of the model, $N$;
    \item the metallicity of the M31 halo contamination, $\FeH_\mathrm{halo}$;
    \item the fractional contribution of MW stars to the total local contamination, $\eta$.
\end{itemize}
\noindent The 5-dimensional grid of parameters for which the algorithm determines a likelihood value are listed in Table~\ref{param}. Similarly to \citetalias{Martin2013}, the favored model is the one that maximizes the likelihood and we use the comparison of this maximal likelihood with the likelihood of the best model with $N\simeq0$ to determine the significance $S$ of a detection.

\begin{figure}
\begin{center}
\includegraphics[width=\hsize]{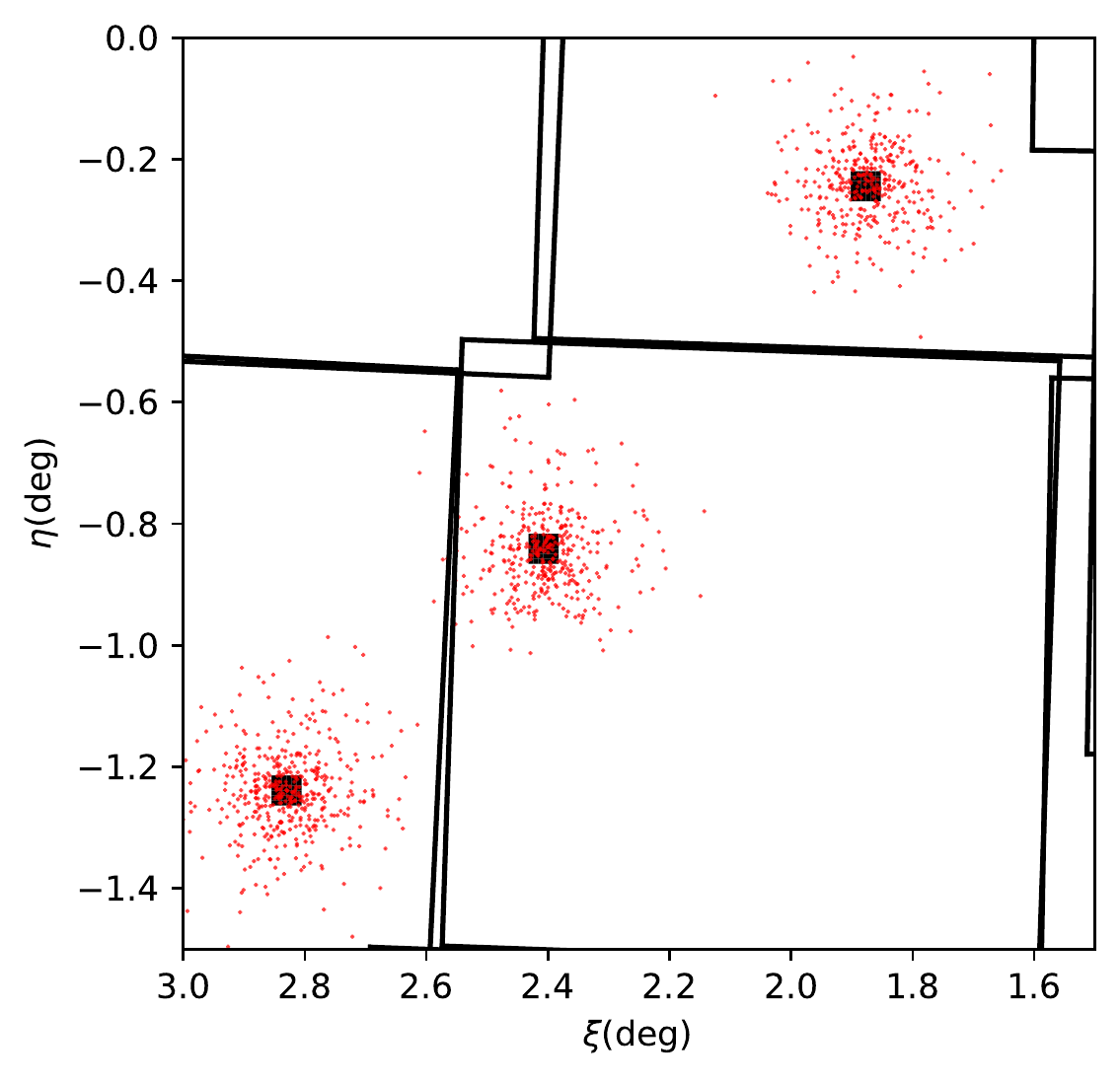}
\caption{\label{mask} Example of the distribution of artificial dwarf galaxies stars in the survey (red points). The large polygons corresponds to the limits of the PAndAS fields in this region. The dark squares each correspond to the $7\times7$ test locations for the search algorithm, and are chosen to be around the known location of an artificial galaxy's center.}
\end{center}
\end{figure}

Since we know the location of the artificial dwarf galaxies ingested in the survey catalogue, we can significantly save on computing time by only running the search algorithm near these locations. Therefore, we only test for the significance of a dwarf galaxy detection within $\pm1.5'$ of the known centers of ingested dwarf galaxies in both the $\xi$ and $\eta$ directions (or $\pm3$ steps on the spatial grid over which the algorithm is run)\footnote{We tested on the field 81 that the recovery fractions are similar when the significance calculations are made within a distance of $\pm2.5'$ and  $\pm1.5'$ of the artificial galaxy center.}. The corresponding locations are highlighted in Figure~\ref{mask} for a run with artificial galaxies that have $r_\mathrm{h}=1\kpc$.

\subsubsection{Choice of the detection threshold}

\begin{figure*}
\begin{center}
\includegraphics[width=\hsize]{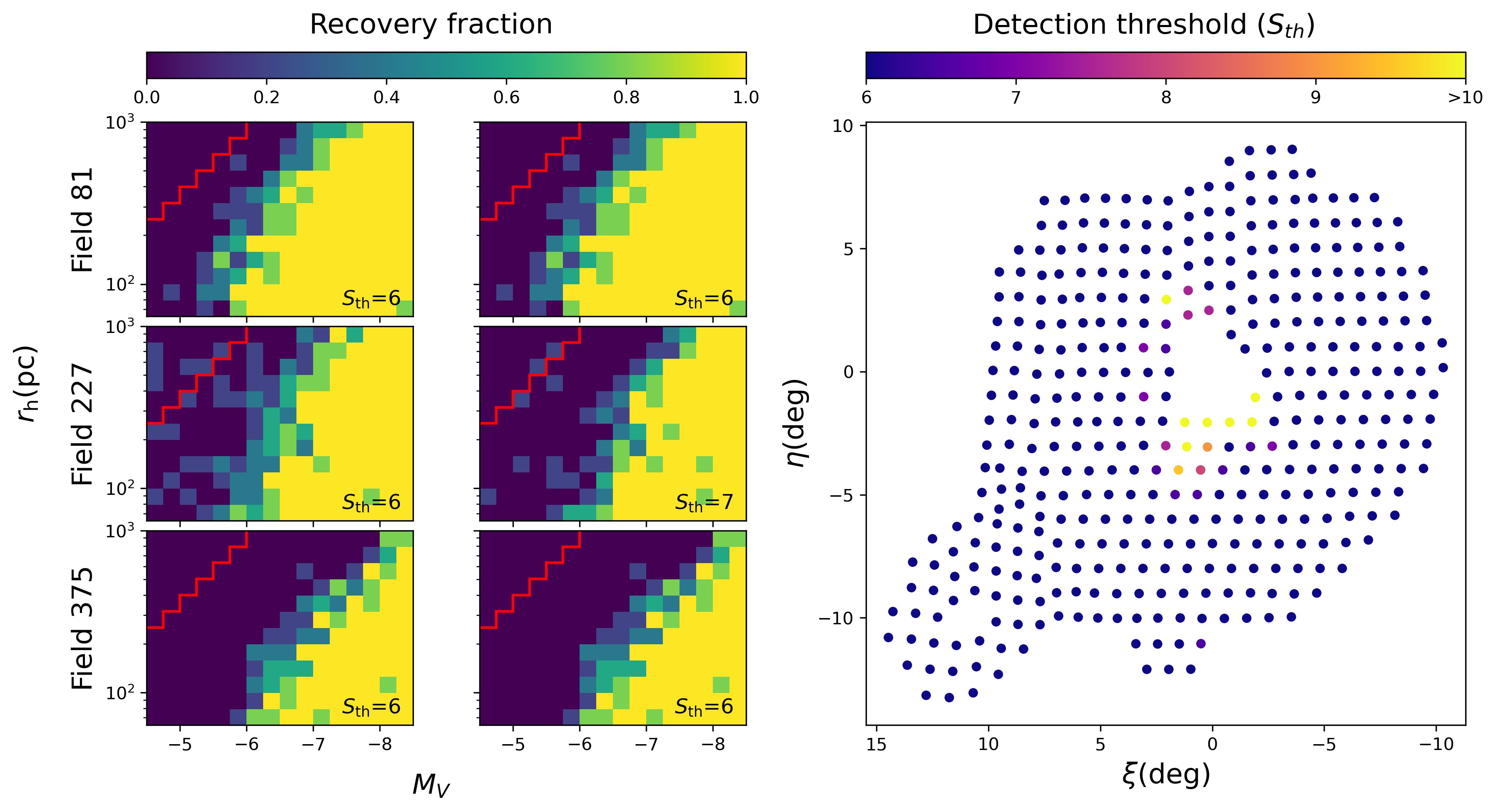}
\caption{\label{seuildetection} \emph{Left:} Recovery fractions for 3 fields of the PAndAS survey, both with the common detection threshold value, $S_\mathrm{th}$=6, and with the updated threshold value determined to minimize the number of false positive in the region of large and faint dwarf galaxies that is highlighted in red. \emph{Right:} Detection thresholds that are used for all considered PAndAS fields. Only fields with a high level of M31 stellar halo contamination have a higher $S_\mathrm{th}$ as the search algorithm struggles to always distinguish the structures of M31 and dwarf galaxies.}
\end{center}
\end{figure*}

The search algorithm outputs the significance $S$ of there being a dwarf galaxy at all tested locations but we still need to determine what constitutes a detection. If the threshold we consider for a detection is too low, it will lead to the detection of systems that would be too faint to be reliably classified as dwarf galaxies in PAndAS and it will not be possible to use the resulting detection limits to understand the M31 satellite system as seen by PAndAS. 
The threshold is derived from the distribution of significances determined by \citetalias{Martin2013} for an area with low M31 and MW contamination near And XI-XIII (see their Figure~9). To determine expectations of detections stemming from noise in the data, we linearly fit the low significance tail of the significance distribution and choose as the significance threshold $S_\mathrm{th}$ the value of $S$ that corresponds to less than one detection in this area of $\sim9$\,deg$^2$. This yields $S_\mathrm{th}=6$. It is coherent with the significance from the dwarf galaxy with the lowest value of $S$ that is unambiguously detected, And~XXVI, for which the algorithm determines $S_\mathrm{th}=5.9$ \citepalias{Martin2013}.

The dwarf galaxy recovery fractions for 3 representative fields are shown in Figure~\ref{seuildetection} for this threshold. These three fields are representative of an outer M31 halo region (field~81), a region contaminated by dense M31 stellar streams (field~227), and a region heavily impacted by the MW foreground contamination (field~375), as highlighted in Figure~\ref{localdensity}. We will discuss the recovery fractions in more detail in the next sub-section, but, at this stage, we wish to focus on the fact that, overall, they behave as expected with faint/compact dwarf galaxies recovered at a high fraction and larger and/or fainter systems showing lower detection rates. However, we note that for field~227 the recovery fractions are non-zero for systems that the search algorithm has no hope of detecting in PAndAS as they are too faint and too extended (region highlighted in red in the figure). This behavior is typical of fields that are heavily contaminated by M31 stellar structures. In these fields, the M31 streams have a color-magnitude distribution that is similar to what the algorithm searches for in a dwarf galaxy, which biases upwards the values of $S$. This was already shown by \citetalias{Martin2013} and it forces us to increase the detection threshold for those fields to avoid false detections and unrealistically high recovery fractions. In practice, we increase $S_\mathrm{th}$ by steps of 0.5 until there are only at most 2 galaxies detected in the region of faint and extended dwarf galaxies highlighted in red in the top-left corner of the recovery-fraction panels of Figure~\ref{seuildetection}. Our aim is to ensure a detection threshold that leads to only a very small number of false-positives, but to not have a threshold that is so conservative that only bright dwarf galaxies are recovered. The resulting recovery rate for field~227 is shown in the second column of panels of the figure, along with the map of final values of $S_\mathrm{th}$ throughout the survey in the right-hand panel. As expected, the nominal detection threshold value, $S_\mathrm{th}=6$, is used for most fields and only fields close to M31 and significantly affected by M31 stellar halo structures require higher detection threshold values.

\subsection{The recovery-fraction model}\label{model}

\begin{figure*}
\begin{center}
\includegraphics[width=\hsize]{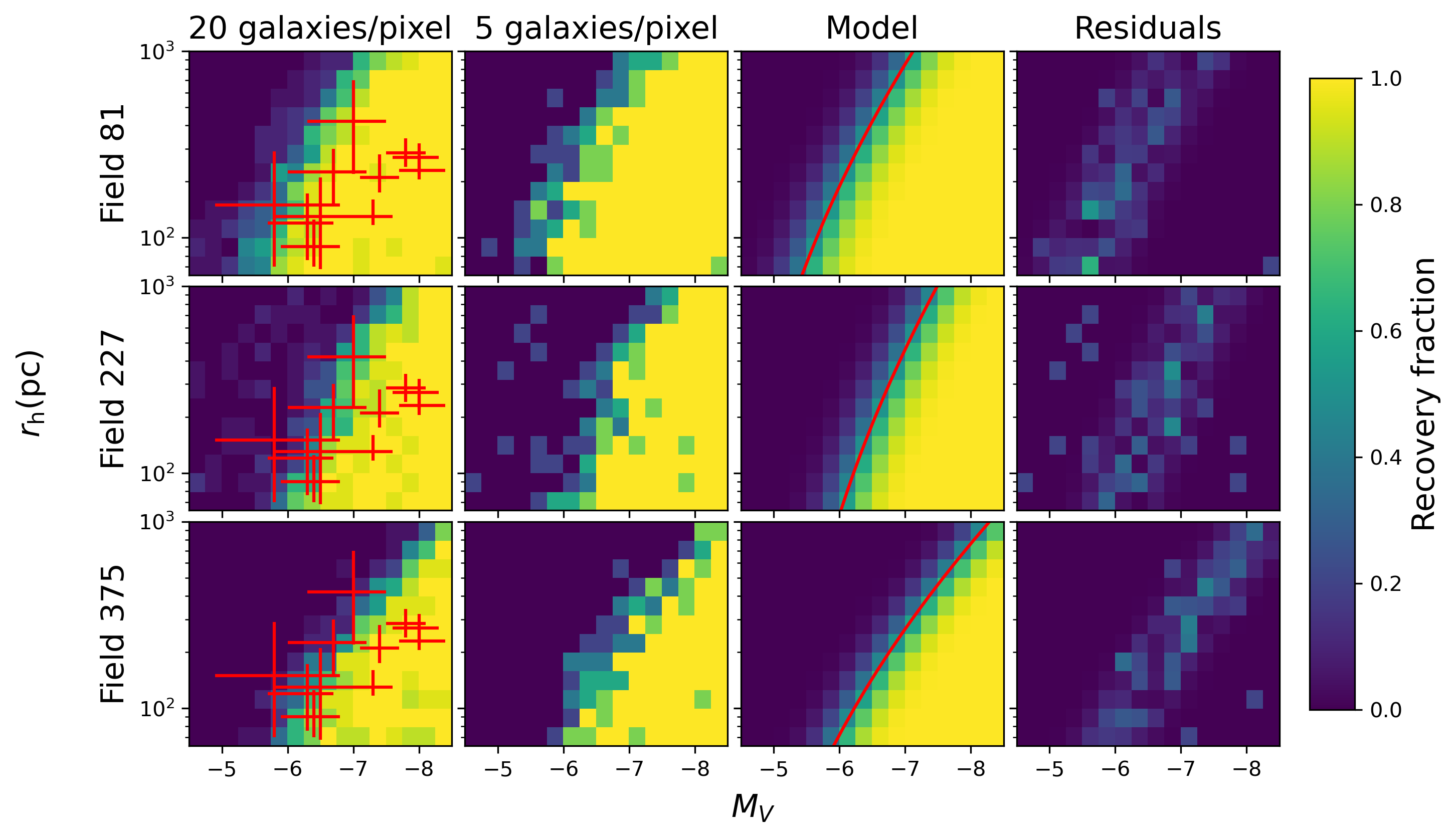}
\caption{\label{DL_model_residu}  Recovery fractions for fields~81, 227, and 375. The first column of panels presents the recovery fractions when simulating 20 galaxies per field and per ($M_{V},\log_{10}(r_\mathrm{h})$) pixel. The red dots correspond to known dwarf galaxies around M31. These straddle the transition region between no recovery and full recovery, as expected. The second column of panels presents the recovery fractions for only 5 galaxies per field and per ($M_{V},\log_{10}(r_\mathrm{h})$) pixel. The third column shows the modeled recovery fractions. In these panels, the red line represents the 50-percent detection threshold. The fourth column displays the residuals that remain small, confirming that our model reproduces the recovery fractions well.}
\end{center}
\end{figure*}

\begin{figure*}
\begin{center}
\includegraphics[width=\hsize]{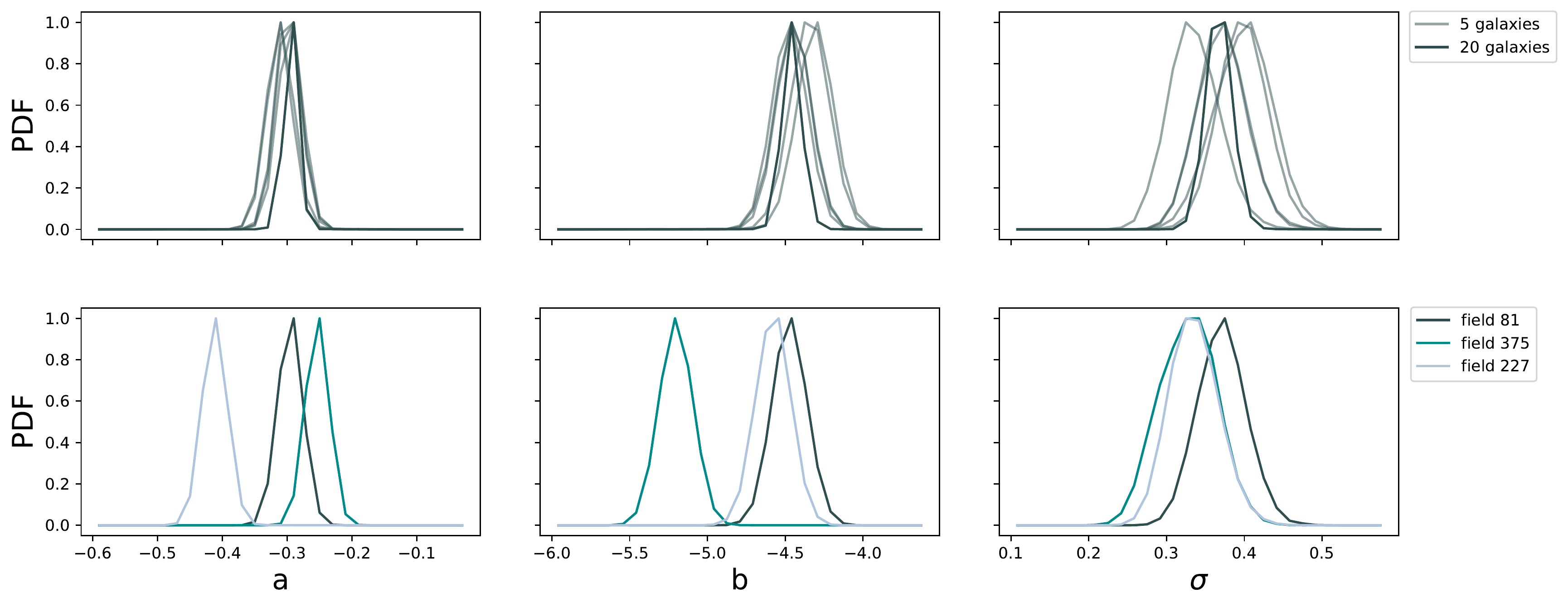}
\caption{\label{PDFs}The top panels show the probability distribution functions for each parameter of the model for the case with 20 and 5 artificial dwarf galaxies per field and per ($M_{V},\log_{10}(r_\mathrm{h})$) pixel for the field 81. The bottom panels show the probability distribution functions for the case of 5 galaxies and for different fields. While using only 5 galaxies per pixel produces wider PDFs, as expected, these show no systematic bias compared to the PDFs resulting from the 20-galaxy case.}
\end{center}
\end{figure*}

The large amount of computing time required to determine the recovery fractions of artificial dwarf galaxies limits our calculations to only 5 simulated dwarf galaxies per field and per magnitude-size pixel, which leads to somewhat noisy recovery fractions. In this sub-section, we aim to build an analytical recovery-fraction model for each field to bypass these limitations.

The left-most column of panels in Figure~\ref{DL_model_residu} shows the outcome of the algorithm for the same three fields~81, 227, and 375 already discussed above. In this case, we run the algorithm with 20 galaxies per ($M_{V},\log_{10}(r_\mathrm{h})$) pixel and per field, to validate that the model we build on the regular simulations with only 5 dwarf galaxies per pixel are representative. In these panels, we also show the properties of known M31 dwarf galaxies located in the PAndAS survey (\citealt{Martin2016}). As expected, these known galaxies are located in the area of high recovery fractions but also cover the area of transition, in agreement with the results of the search algorithm \citepalias{Martin2013} and our choice for the detection threshold values. For comparison, the second column of panels displays the recovery fractions of dwarf galaxies with our regular set up of 5~dwarf galaxies per pixel. It is reassuring that, even though the recovery fractions are here more noisy, as expected with fewer artificial galaxies, they do look similar to the more accurate results with 20~galaxies per pixel. We will provide a quantitative comparison below.

The recovery fractions behave as expected: high surface-brightness systems, i.e. compact and/or bright dwarf galaxies in the bottom-right half of the plots, are recovered at high efficiency while more extended and/or fainter systems are progressively missed by the search algorithm. The field contamination (mainly from the M31 stellar halo in field 227 or the MW in field 375) has an impact on the recovery fractions and, in general, degrades them. We note that, in the case of a high level of contamination from M31 stellar halo structures, the algorithm recovers a small number of large and faint dwarf galaxies below the transition region. As mentioned above, these are false detections and an artifact of the search algorithm that is sometimes struggling to discriminate contaminating M31 stream stars from the dwarf galaxy stars at the same location of the CMD because they all follow similar RGB tracks. However, this effect remains small since, as mentioned above, we adapt the threshold limit to minimize the presence of these false detections. This effect is not present for fields heavily contaminated by foreground MW stars (field 375 in Figure~\ref{DL_model_residu}) since these contaminating stars have a color-magnitude distribution that is quite different from that of the M31 (artificial) dwarf galaxies \citepalias{Martin2013}. In this case, the added contamination leads to recovery fractions that are shifted towards brighter and/or more compact systems.

We fit an analytical model to the recovery fractions to overcome the noise and the false detections. Given the shape of the recovery-fraction distributions, our analytical model is built around a quadratic transition region in the $M_V$--$\log_{10}(r_\mathrm{h})$ plane. We allow for the parameters of this quadratic transition to vary from field to field and we parametrize the width of the smooth transition between regions of full to no recovery.

We define the model for the recovery fraction of artificial dwarf galaxies as

\begin{equation}
\begin{aligned}
\label{model}
f(M_V,\log_{10}(r_\mathrm{h}))&=F\left(\frac{M_V-M_{V,\mathrm{lim}}}{\sigma}\right), \\
\textrm{with } M_{V,\mathrm{lim}}&=a\log_{10}(r_\mathrm{h})^2+b.
\end{aligned}
\end{equation}

\noindent Here, a and b do not carry any significant physical meaning and are just used to parametrize the space, $\sigma$ is the width of the transition, and $F$ is the complementary error function, defined as

\begin{equation}
    F(x)=\frac{1}{2} \erfc\left(\frac{x}{\sqrt{2}} \right).
\end{equation}

For each field, we fit the model to the recovery-fraction distribution in the $M_{V}$--$\log_{10}(r_\mathrm{h})$ plane by determining the likelihood of the data given the model thanks to a Metropolis–Hastings method. The variables a, b and $\sigma$ are only used to parametrize the model, the important result is the threshold in surface brightness (which we will use for further analyses). Then, the best set of parameters is obtained by calculating the binomial likelihood of the number of recovered dwarf galaxies compared to the model. For the three representative fields and the case with 5 artificial galaxies per pixel, the resulting best fit models are displayed in the third column of panels in Figure~\ref{DL_model_residu}, with the residuals shown in the fourth column. The models provide a good representation of the recovery fractions and the residuals remain small.

To confirm that we do not introduce any bias by simulating only 5 galaxies per ($M_{V},\log_{10}(r_\mathrm{h})$) pixel, in Figure~\ref{PDFs}, we compare the marginalized probability distribution functions (PDFs) obtained for the three parameters in the fields 81 when using 20 (dashed lines) and sub-samples of 5 (full lines) simulated galaxies per pixel. Even though they are, as expected wider, we find no systematic bias in the PDFs obtained when using only 5 dwarf galaxies per pixel. Therefore, we go ahead and determine the recovery fractions with 5 artificial galaxies per pixel and per field as it already represents more than 90 000 CPU hours on the Strasbourg University High Performance Computing center for the simulations over the $\sim400$ PAndAS fields.

\section{Results} \label{Results}

\subsection{Dwarf galaxy recovery fractions over the PAndAS survey}
\begin{figure*}
\begin{center}
\includegraphics[width=\hsize]{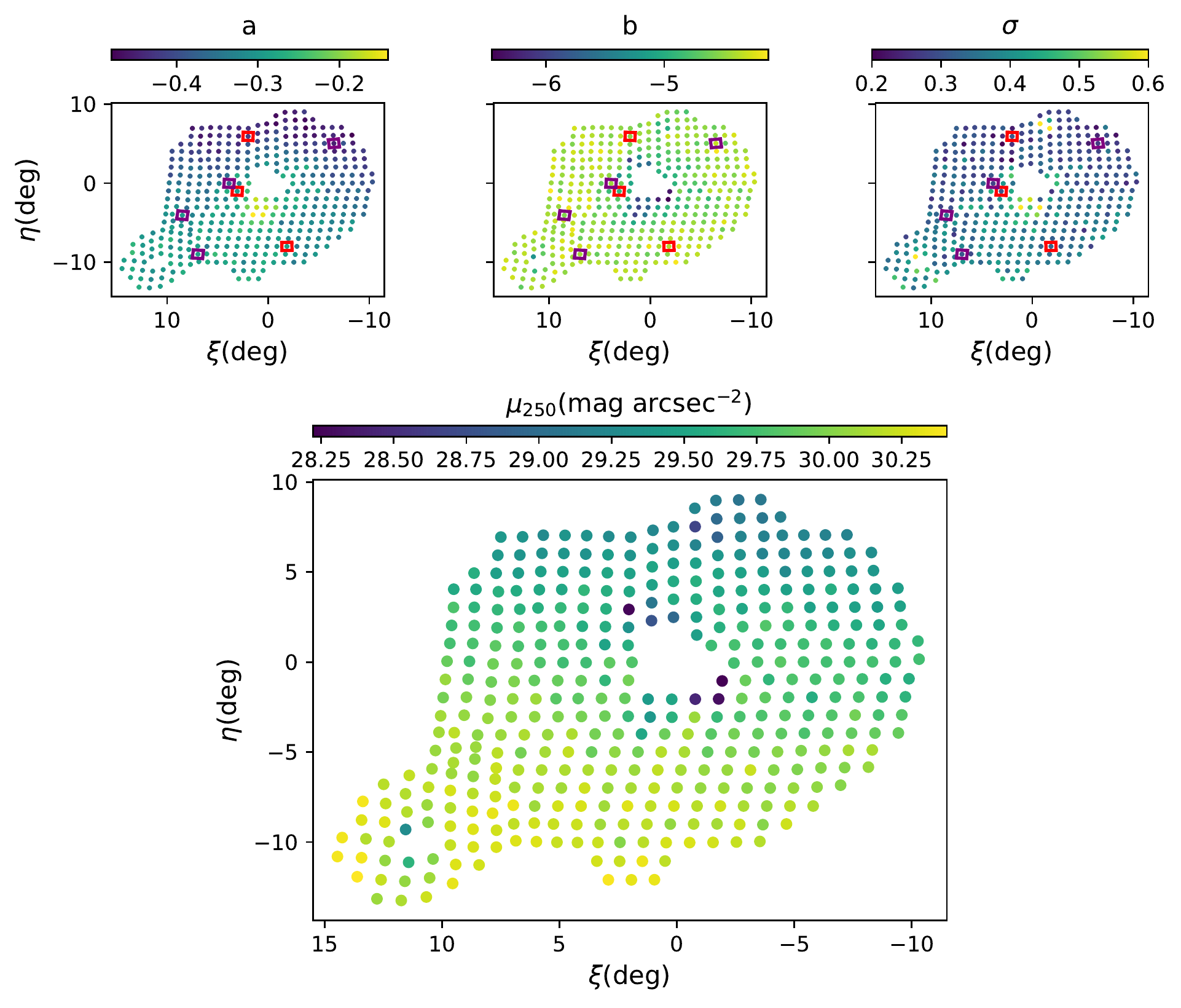}
\caption{\label{carteparam} Values of the favored model parameters, a, b and $\sigma$, for each field in the PAndAS survey (top three panels). The bottom panel translates those into the surface brightness within the half-light radius 50\% detection limit for a dwarf galaxy with $r_\mathrm{h}= 250\pc$ (see text for more details). The fields highlighted in red are those chosen to be representative of the different regions of the survey while those represented in purple are the additional fields use to test the impact of the distance on the recovery fractions. The surface brightness thresholds varie depending on the position: in regions with low contamination fainter galaxies can be detected compared to regions with significant M31 or MW contamination. }
\end{center}
\end{figure*}

Figure~\ref{carteparam} summarizes the recovery fractions of all studied PAndAS fields after ingesting more than 350,000 artificial dwarf galaxies in the stellar catalogue. The top three panels of the figure show, for each field, the values of the three model parameters over the survey. There are some clear changes to the values of parameters a and b that can be tracked to changes in the properties of the survey: the north-south gradient is linked to the increased MW foreground contamination towards the north and the presence of the Giant Stream south of M31 ($0\lta\xi\lta5\deg$ and $-5\lta\eta\lta0\deg$) clearly impacts the values of a. On the other hand, the speed of the transition, parametrized by $\sigma$, remains fairly constant over the survey.

The slope of the transition region in the $\log_{10}(r_\mathrm{h})$--$M_V$ plane does not map lines of constant surface brightness. While this is common \cite[e.g.][]{Koposov2008,Walsh2009,Drlica2020-1} and is likely related to the complex nature of the data and of the search algorithm, it can make it difficult to interpret our results in term of surface brightness limits. We therefore show, in the bottom panel of Figure~\ref{carteparam}, the corresponding surface brightness, $\mu_{250}$, within the half-light radius of a system for a fixed $r_\mathrm{h}= 250\pc$. Most fields show $29<\mu_{250}<30$ mag/arcsec$^2$, with a clear impact from the MW contamination that is significantly more important on the northern side of the survey that reaches Galactic latitude $b~-12\deg$. One can also note the impact of the M31 stellar halo contamination in the regions nearest M31. The necessity to increase the detection threshold $S_\mathrm{th}$ to avoid false-positives is partly responsible (Figure~\ref{DL_model_residu}) but the lower surface brightness limit is also linked to a lower contrast because of a higher density of contaminating M31 stars. The model parameters for all fields are listed in Table~\ref{Model_parameters}.

\subsection{Impact of fixed artificial dwarf galaxy parameters}\label{Results2}

\begin{figure*}
\begin{center}
\includegraphics[width=\hsize]{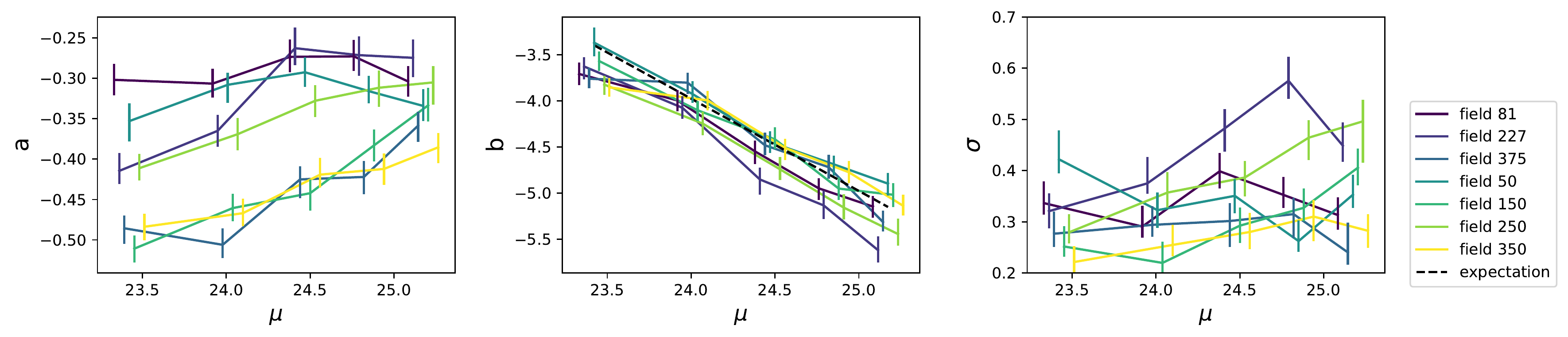}
\caption{\label{paramdistance}  Changes to the best parameters of the recovery fraction model as a function of the distance modulus to the dwarf galaxy. Each point is slightly shifted in distance, in order to make the graph more legible. The black dashed line represents the theoretical change of $M_V$ in function of the distance modulus for the field 50. Parameters a and b smoothly change with distance, which allows us to easily model the influence of the distance on the recovery fractions. }
\end{center}
\end{figure*}
\begin{figure*}
\begin{center}
\includegraphics[width=\hsize]{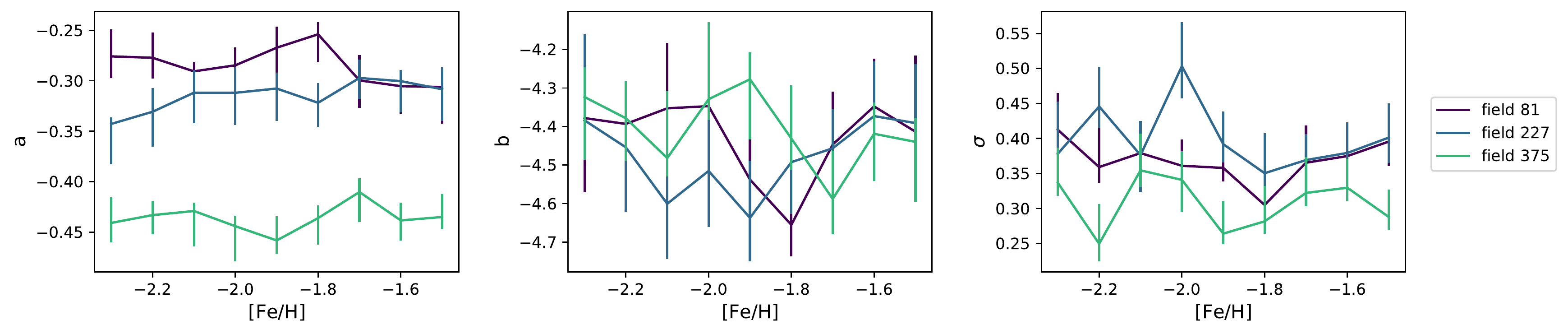}
\caption{\label{parammetallicite}  Changes to the best parameters of the recovery fraction model as a function of a dwarf galaxy's metallicity. Variation to the metallicity within this metal-poor range do not significantly change the parameters of the model. }
\end{center}
\end{figure*}

As mentioned in Section~\ref{Methods}, we fixed some of the supposedly less impactful parameters of the artificial dwarf galaxies to save on the computing time that is already quite large. Here, we test the impact of these parameters on the recovery fraction model parameters for seven representative fields of the survey. Those are chosen to be dispersed throughout the survey to represent different types of contamination. 

To determine the impact of the distance to the artificial dwarf galaxies, we conduct a set of simulations with 5 galaxies/pixel and distances that vary between $\pm300\kpc$ of the distance to M31 that is assumed for the main run. The resulting variations in the recovery model parameters are shown in the Figure~\ref{paramdistance} for steps of $150\kpc$. One can expect that for a closer/further distance, the recovery fractions shown in Figure~\ref{DL_model_residu} would simply shift to the top-left/bottom-right. This is indeed the main effect that we seen in the left-hand panel of Figure~\ref{paramdistance}, with b typically shifting by the distance modulus change that corresponds to the distance change. We note, however that changes in the distance to the artificial dwarf galaxies also have some impact on a of the recovery fraction model (middle panel). While the reason for this correlation is not readily evident, it may be related to the algorithm needing a minimum number of stars in the M31 CMD box to yield a reliable detection, with a smaller impact from the size of the system in this regime. Such an effect would tend to make the slope of the transition region of the recovery fractions steeper as the number of stars in the CMD box becomes smaller. It could explain why more distant artificial galaxies, which have fewer stars in the magnitude-limited CMD box, yield higher values of a.

Irrespective of the reason for the changes of a and b with distance, these anticorrelations are regular and can easily be modeled. This has the benefit of preventing numerous additional simulations while still taking into account that dwarf galaxies significantly in front or behind M31 have somewhat different recovery fractions. We therefore fit a linear model to the values of a and b shown in Figure~\ref{paramdistance}, using the data from all fields together, but allowing for different intercepts for each field. We define as a$_0$ and b$_0$ the intercepts (their values are accessible in the appendix) and $\mu$ the distance modulus. The resulting fits yield

\begin{equation}
\begin{aligned}
a&=0.06\mu+a_0, \mathrm{and} \\
b&=-0.89\mu+b_0.
\end{aligned}
\end{equation}

The third parameter of the model, $\sigma$, which represents the width of the transition region of the recovery fractions in the $\log_{10}(r_\mathrm{h})$--$M_V$ space, shows no significant changes with distance. We therefore assume that it is not impacted by changes in the distance to the dwarf galaxies.

In Section~\ref{Methods}, we also fixed the metallicity of the metal-poor artificial dwarf galaxies to $\FeH=-1.7$ as the color of the RGB only shifts slightly in color in this metal-poor regime. The impact of this choice of metallicity is explored by determining the recovery fraction for artificial galaxies in the range $-2.3<\FeH<-1.5$, with steps of 0.1\,dex, for fields 81, 227, and 375. The results are statistically similar over the metallicity range, as shown in Figure~\ref{parammetallicite}. We conclude that the impact of $\FeH$ on the recovery fraction is negligible and that our assumption of $\FeH=-1.7$ does not affect our results.

\subsection{Using the recovery fractions}

To facilitate the use of the previous results and, for instance, easily apply the PAndAS recovery fraction to sets of simulated dwarf galaxy satellite system, we make public a python module that is accessible at \url{https://github.com/dolivadolinsky/Recovery_dwarf_galaxy_M31}. Given the cartesian coordinates of a dwarf galaxy around its host, its half-light radius, and its V-band magnitude, this module returns the probability that such a dwarf galaxy would have been discovered in PAndAS. The module first determines the fields in which the tested dwarf galaxy would be located or if it is outside the PAndAS footprint. If there is an overlap of fields, the parameters values for the galaxy are the ones of the deepest photometry. Then, the relations described in Section~\ref{Results2} are used to determine the values of a and b at the galaxy's distance in order to retrieve the recovery fractions. 

\section{Discussion and conclusion}  \label{Discussion}
In this paper, we determined the recovery fraction of Andromeda's dwarf galaxy based on the algorithm of \citetalias{Martin2013} applied to the photometric data from the PAndAS survey. We repeatedly simulated artificial dwarf galaxies, blindly added them to the PAndAS photometric catalogue, and checked the significant of their recovery by running the search algorithm. We determined and modeled the recovery fractions of satellites on a field-to-field basis for the 390 $1\deg\times1\deg$ fields of the survey and as a function of a dwarf galaxy's magnitude, half-light radius, and heliocentric distance. We publish the resulting recovery fraction as a python module that provides the probability of observing a dwarf galaxy of a given magnitude, half-light radius, and distance in the PAndAS survey.
 
As expected, we found that the recovery fractions are highly sensitive to the size and luminosity of a dwarf galaxy but also to its position due to the significant changes in the level of MW and M31 stellar halo contamination throughout the PAndAS footprint. Constant recovery fractions do not map constant surface brightness levels, likely because of the complex nature of the data and of the search algorithm. For galaxies with $r_\mathrm{h} =250\pc$, the surface brightness threshold that corresponds to a 50-percent detection rate mainly varies between $\sim29.0$\,mag arcsec$^{-2}$ and 30.0\,mag arcsec$^{-2}$, with a median value of 29.8. These results are in good agreement with the surface brightness of genuine dwarf galaxies discovered in PAndAS \citep{Martin2016}. Depending on the field of the survey these limits can be as low as 28.0\,mag arcsec$^{-2}$ or as high and 30.5\,mag arcsec$^{-2}$. The completeness of the survey also depends on the distance to a dwarf galaxy but this variation of the surface brightness threshold roughly maps expectations from changes to the distance modulus.

Our results bear similarities with studies based on other data sets but using the same methodology of inserting stars from artificial dwarf galaxies at the photometric catalogue level. Even if the given surface brightness threshold values are not always comparable, we can calculate the $\mu_{250}$ values based on the results from other studies. \citet{Koposov2008} and \citet{Walsh2009} derive $\mu_{250}\sim28$ mag arcsec$^{-2}$ at the distance of M31, but for the significantly shallower SDSS data and without considering the specific of the region around M31 that becomes significantly contaminated by foreground MW stars. Recently, \citet{Drlica2020-1} determined dwarf galaxy recovery fractions for a large part of the sky from both the Pan-STARRS1 and DES photometric data. The surface brightness thresholds obtained for the two surveys in this study are 28 and 30 mag arcsec$^{-2}$, respectively. It may seem surprising that these results based on the DES are similar to those we obtain for the deeper PAndAS data. First, we note that the PAndAS footprint (latitude $\lvert$b$\lvert\sim22\deg$) is more contaminated by MW and M31 stars than a typical DES field (latitude $\lvert$b$\lvert\geq30\deg$). In addition, the surface brightness limits are sensitive to the distance and the depth of surveys as well as the algorithm use to search for dwarf galaxies, those can be an explanation for the similarity between the DES thresholds and the ones obtained in this work. Finally, the detection threshold ($S_\mathrm{th}$ in our work) used by \citet{Drlica2020-1} to determine if an artificial dwarf galaxy is recovered is particularly low ($S_\mathrm{th}=4.9$). In fact, it is lower than the value $S_\mathrm{th}=6.0$ used by \citep{Drlica2015} to detect candidate satellite like Tucana V or Cetus II that were not confirmed as well-defined, localized stellar overdensities from deeper photometry \citep{Conn2018-1, Conn2018-2}. A particularly aggressive detection threshold would naturally lead to deeper detection limits, but these would be over-estimated compared to the detection limits of genuine dwarf galaxies as the low threshold would lead to false positives in the list of candidate detections.

It is also interesting to compare our results with those of \cite{Garling2021} for the more distant galaxies NGC~3077 and NGC~4214, located more than 3 Mpc away. While the data of the LBT-SONG survey used by these authors are comparable or deeper than the PAndAS data, the larger distance forces them to ingest artificial dwarf-galaxy stars directly in the images to properly account for crowding that becomes an issue because of the smaller angular size of the systems. Their analysis yields $\mu_{250}\sim28$ mag arcsec$^{-2}$, towards the brighter end of the detection limits we determined from M31 and shows one of the limitations of ground-based surveys beyond the Local Group.

Finally, we compare our results with the ones of \cite{Huxor2014}, who determined the globular cluster completeness of the PAndAS survey. They found a 50\% detection surface brightness inside the half-light radius threshold of $\sim$ 26\,mag/arcsec$^2$. Compared to our results, the brighter limit can be explain by the fact that they looked for more compact object and that the search for globular cluster was performed through visual inspection of the images. 

The work presented here complements our build up of a thorough understanding of the dwarf galaxy satellite system of M31 \citep{Martin2013,Martin2016} by clearly expressing the limits to our dwarf galaxy searches within the PAndAS survey. Our results can be seen as an intermediate step towards the dwarf galaxy searches that will be possible within $\sim1$\,Mpc with the next generation of ground-based photometric surveys like the Legacy Survey of Space and Time \citep[LSST;][]{Ivezic2019}. More immediately, we aim to use these recovery fractions to infer the underlying properties of the M31 dwarf galaxy satellite system (number of dwarf galaxies, distribution, size-luminosity relation, etc.) to provide constraints on dwarf galaxy formation and evolution in a cosmological context (Doliva-Dolinsky et al., in preparation).

\acknowledgements
Based on observations obtained with MegaPrime/ MegaCam, a joint project of CFHT and CEA/DAPNIA, at the Canada-France-Hawaii Telescope (CFHT), which is operated by the National Research Council (NRC) of Canada, the Institut National des Science de l'Univers of the Centre National de la Recherche Scientifique (CNRS) of France, and the University of Hawaii.
The authors would like to acknowledge the High Performance Computing Center of the University of Strasbourg for supporting this work by providing scientific support and access to computing resources. Part of the computing resources were funded by the Equipex Equip@Meso project (Programme Investissements d'Avenir) and the CPER Alsacalcul/Big Data. 
NFM, RI, and ZY gratefully acknowledge support from the French National Research Agency (ANR) funded project ``Pristine'' (ANR-18-CE31-0017) and from the European Research Council (ERC) under the European Unions Horizon 2020 research and innovation programme (grant agreement No. 834148).
G.T. acknowledge support from the Agencia Estatal de Investigaci\'on (AEI) of the Ministerio de Ciencia e Innovaci\'on (MCINN) under grant with reference (FJC2018-037323-I).


\begin{appendix}
\section{Model parameters}

\label{Model_parameters}
\begingroup
\centering
\def\arraystretch{1}
           \begin{longtable}{lccccc}
           \hline
Field & a & b& $\sigma$ &a$_0$ &b$_0$ \\
\hline
\endhead
1 & -0.27 & -4.69 & 0.48 & -1.74 & 14.89 \\
2 & -0.31 & -4.40 & 0.34 & -1.78 & 15.18 \\
3 & -0.29 & -4.41 & 0.41 & -1.76 & 15.17 \\
4 & -0.28 & -4.49 & 0.38 & -1.75 & 15.09 \\
5 & -0.32 & -4.42 & 0.38 & -1.79 & 15.16 \\
6 & -0.28 & -4.30 & 0.48 & -1.75 & 15.28 \\
7 & -0.27 & -4.44 & 0.51 & -1.74 & 15.14 \\
8 & -0.31 & -4.44 & 0.37 & -1.78 & 15.14 \\
9 & -0.26 & -4.44 & 0.46 & -1.73 & 15.14 \\
10 & -0.26 & -4.5 & 0.43 & -1.73 & 15.08 \\
11 & -0.28 & -4.37 & 0.35 & -1.75 & 15.21 \\
12 & -0.29 & -4.28 & 0.36 & -1.76 & 15.30 \\
13 & -0.31 & -4.48 & 0.40 & -1.78 & 15.10 \\
14 & -0.31 & -4.90 & 0.51 & -1.78 & 14.68 \\
15 & -0.27 & -4.40 & 0.36 & -1.74 & 15.18 \\
16 & -0.31 & -4.24 & 0.41 & -1.78 & 15.34 \\
17 & -0.27 & -4.51 & 0.40 & -1.74 & 15.07 \\
18 & -0.30 & -4.58 & 0.47 & -1.77 & 15.00 \\
19 & -0.31 & -4.28 & 0.42 & -1.78 & 15.30 \\
20 & -0.30 & -4.37 & 0.34 & -1.77 & 15.21 \\
21 & -0.28 & -4.36 & 0.42 & -1.75 & 15.22 \\
22 & -0.28 & -4.52 & 0.46 & -1.75 & 15.06 \\
23 & -0.29 & -4.29 & 0.34 & -1.76 & 15.29 \\
24 & -0.31 & -4.38 & 0.42 & -1.78 & 15.20 \\
25 & -0.30 & -4.44 & 0.36 & -1.77 & 15.14 \\
27 & -0.28 & -4.43 & 0.34 & -1.75 & 15.15 \\
28 & -0.27 & -4.55 & 0.33 & -1.74 & 15.03 \\
29 & -0.28 & -4.42 & 0.41 & -1.75 & 15.16 \\
31 & -0.29 & -4.36 & 0.41 & -1.76 & 15.22 \\
32 & -0.27 & -4.47 & 0.39 & -1.74 & 15.11 \\
33 & -0.30 & -4.37 & 0.33 & -1.77 & 15.21 \\
34 & -0.28 & -4.48 & 0.39 & -1.75 & 15.10 \\
35 & -0.26 & -4.59 & 0.42 & -1.73 & 14.99 \\
36 & -0.33 & -4.41 & 0.38 & -1.8 & 15.17 \\
37 & -0.27 & -4.50 & 0.37 & -1.74 & 15.08 \\
38 & -0.34 & -4.15 & 0.39 & -1.81 & 15.43 \\
39 & -0.30 & -4.33 & 0.36 & -1.77 & 15.25 \\
40 & -0.29 & -4.37 & 0.36 & -1.76 & 15.21 \\
41 & -0.30 & -4.43 & 0.37 & -1.77 & 15.15 \\
42 & -0.31 & -4.35 & 0.37 & -1.78 & 15.23 \\
43 & -0.31 & -5.26 & 1.13 & -1.78 & 14.32 \\
44 & -0.25 & -4.58 & 0.39 & -1.72 & 15.00 \\
45 & -0.29 & -4.35 & 0.37 & -1.76 & 15.23 \\
46 & -0.31 & -4.24 & 0.28 & -1.78 & 15.34 \\
47 & -0.28 & -4.46 & 0.36 & -1.75 & 15.12 \\
48 & -0.28 & -4.46 & 0.38 & -1.75 & 15.12 \\
49 & -0.28 & -4.69 & 0.44 & -1.75 & 14.89 \\
50 & -0.29 & -4.45 & 0.31 & -1.76 & 15.13 \\
51 & -0.28 & -4.46 & 0.29 & -1.75 & 15.12 \\
52 & -0.25 & -4.66 & 0.41 & -1.72 & 14.92 \\
53 & -0.27 & -4.53 & 0.37 & -1.74 & 15.05 \\
54 & -0.30 & -4.35 & 0.39 & -1.77 & 15.23 \\
55 & -0.33 & -4.41 & 0.34 & -1.80 & 15.17 \\
56 & -0.31 & -4.44 & 0.36 & -1.78 & 15.14 \\
57 & -0.30 & -4.36 & 0.39 & -1.77 & 15.22 \\
58 & -0.30 & -4.43 & 0.43 & -1.77 & 15.15 \\
59 & -0.28 & -4.53 & 0.38 & -1.75 & 15.05 \\
60 & -0.29 & -4.46 & 0.41 & -1.76 & 15.12 \\
61 & -0.29 & -4.54 & 0.34 & -1.76 & 15.04 \\
62 & -0.28 & -4.65 & 0.39 & -1.75 & 14.93 \\
63 & -0.27 & -4.61 & 0.38 & -1.74 & 14.97 \\
64 & -0.30 & -4.27 & 0.42 & -1.77 & 15.31 \\
65 & -0.26 & -4.45 & 0.35 & -1.73 & 15.13 \\
66 & -0.29 & -4.37 & 0.40 & -1.76 & 15.21 \\
67 & -0.30 & -4.38 & 0.38 & -1.77 & 15.20 \\
68 & -0.32 & -4.31 & 0.33 & -1.79 & 15.27 \\
69 & -0.26 & -4.74 & 0.41 & -1.73 & 14.84 \\
70 & -0.29 & -4.31 & 0.38 & -1.76 & 15.27 \\
71 & -0.34 & -4.27 & 0.36 & -1.81 & 15.31 \\
72 & -0.30 & -4.45 & 0.28 & -1.77 & 15.13 \\
73 & -0.30 & -4.33 & 0.40 & -1.77 & 15.25 \\
74 & -0.32 & -4.30 & 0.35 & -1.79 & 15.28 \\
75 & -0.31 & -4.26 & 0.39 & -1.78 & 15.32 \\
76 & -0.29 & -4.49 & 0.39 & -1.76 & 15.09 \\
77 & -0.33 & -4.35 & 0.35 & -1.80 & 15.23 \\
78 & -0.29 & -4.55 & 0.38 & -1.76 & 15.03 \\
79 & -0.34 & -4.23 & 0.31 & -1.81 & 15.35 \\
80 & -0.29 & -4.35 & 0.39 & -1.76 & 15.23 \\
81 & -0.29 & -4.4 & 0.33 & -1.76 & 15.18 \\
82 & -0.27 & -4.57 & 0.36 & -1.74 & 15.01 \\
83 & -0.30 & -4.42 & 0.33 & -1.77 & 15.16 \\
84 & -0.32 & -4.21 & 0.36 & -1.79 & 15.37 \\
85 & -0.29 & -4.41 & 0.40 & -1.76 & 15.17 \\
86 & -0.33 & -4.25 & 0.37 & -1.80 & 15.33 \\
87 & -0.30 & -4.42 & 0.28 & -1.77 & 15.16 \\
88 & -0.26 & -4.56 & 0.36 & -1.73 & 15.02 \\
89 & -0.30 & -4.39 & 0.36 & -1.77 & 15.19 \\
90 & -0.27 & -4.65 & 0.38 & -1.74 & 14.93 \\
91 & -0.29 & -4.52 & 0.32 & -1.76 & 15.06 \\
92 & -0.32 & -4.43 & 0.40 & -1.79 & 15.15 \\
93 & -0.32 & -4.41 & 0.37 & -1.79 & 15.17 \\
94 & -0.28 & -4.60 & 0.38 & -1.75 & 14.98 \\
95 & -0.34 & -4.35 & 0.42 & -1.81 & 15.23 \\
96 & -0.32 & -4.47 & 0.32 & -1.79 & 15.11 \\
97 & -0.28 & -4.47 & 0.35 & -1.75 & 15.11 \\
98 & -0.32 & -4.27 & 0.37 & -1.79 & 15.31 \\
99 & -0.33 & -4.39 & 0.33 & -1.80 & 15.19 \\
100 & -0.30 & -4.51 & 0.37 & -1.77 & 15.07 \\
101 & -0.32 & -4.40 & 0.32 & -1.79 & 15.18 \\
102 & -0.30 & -4.47 & 0.37 & -1.77 & 15.11 \\
103 & -0.29 & -4.59 & 0.34 & -1.76 & 14.99 \\
104 & -0.28 & -4.51 & 0.38 & -1.75 & 15.07 \\
105 & -0.31 & -4.42 & 0.37 & -1.78 & 15.16 \\
106 & -0.32 & -4.27 & 0.27 & -1.79 & 15.31 \\
107 & -0.30 & -4.55 & 0.41 & -1.77 & 15.03 \\
108 & -0.31 & -4.45 & 0.36 & -1.78 & 15.13 \\
109 & -0.30 & -4.49 & 0.36 & -1.77 & 15.09 \\
110 & -0.27 & -4.61 & 0.33 & -1.74 & 14.97 \\
111 & -0.33 & -4.19 & 0.31 & -1.80 & 15.39 \\
112 & -0.30 & -4.52 & 0.39 & -1.77 & 15.06 \\
113 & -0.30 & -4.45 & 0.35 & -1.77 & 15.13 \\
114 & -0.29 & -4.65 & 0.38 & -1.76 & 14.93 \\
115 & -0.29 & -4.52 & 0.36 & -1.76 & 15.06 \\
116 & -0.30 & -4.45 & 0.42 & -1.77 & 15.13 \\
117 & -0.33 & -4.43 & 0.26 & -1.80 & 15.15 \\
118 & -0.34 & -4.35 & 0.37 & -1.81 & 15.23 \\
119 & -0.32 & -4.47 & 0.30 & -1.79 & 15.11 \\
120 & -0.28 & -4.57 & 0.40 & -1.75 & 15.01 \\
121 & -0.31 & -4.31 & 0.35 & -1.78 & 15.27 \\
122 & -0.25 & -4.75 & 0.43 & -1.72 & 14.83 \\
123 & -0.32 & -4.41 & 0.37 & -1.79 & 15.17 \\
124 & -0.27 & -4.69 & 0.37 & -1.74 & 14.89 \\
125 & -0.29 & -4.59 & 0.35 & -1.76 & 14.99 \\
126 & -0.28 & -4.51 & 0.34 & -1.75 & 15.07 \\
127 & -0.28 & -4.59 & 0.45 & -1.75 & 14.99 \\
128 & -0.30 & -4.55 & 0.40 & -1.77 & 15.03 \\
129 & -0.32 & -4.41 & 0.31 & -1.79 & 15.17 \\
130 & -0.31 & -4.35 & 0.40 & -1.78 & 15.23 \\
131 & -0.33 & -4.32 & 0.34 & -1.80 & 15.26 \\
132 & -0.29 & -4.70 & 0.32 & -1.76 & 14.88 \\
133 & -0.30 & -4.46 & 0.41 & -1.77 & 15.12 \\
134 & -0.34 & -4.24 & 0.35 & -1.81 & 15.34 \\
135 & -0.30 & -4.48 & 0.36 & -1.77 & 15.10 \\
136 & -0.33 & -4.34 & 0.37 & -1.80 & 15.24 \\
137 & -0.33 & -4.37 & 0.36 & -1.80 & 15.21 \\
138 & -0.29 & -4.44 & 0.41 & -1.76 & 15.14 \\
139 & -0.32 & -4.46 & 0.37 & -1.79 & 15.12 \\
140 & -0.30 & -4.60 & 0.40 & -1.77 & 14.98 \\
141 & -0.28 & -4.80 & 0.39 & -1.75 & 14.78 \\
142 & -0.30 & -4.64 & 0.30 & -1.77 & 14.94 \\
143 & -0.26 & -4.78 & 0.40 & -1.73 & 14.80 \\
144 & -0.34 & -4.37 & 0.34 & -1.81 & 15.21 \\
145 & -0.23 & -5.05 & 0.46 & -1.70 & 14.53 \\
146 & -0.35 & -4.42 & 0.40 & -1.82 & 15.16 \\
147 & -0.23 & -4.83 & 0.45 & -1.70 & 14.75 \\
148 & -0.25 & -4.75 & 0.34 & -1.72 & 14.83 \\
149 & -0.31 & -4.43 & 0.37 & -1.78 & 15.15 \\
150 & -0.30 & -4.56 & 0.36 & -1.77 & 15.02 \\
151 & -0.36 & -4.35 & 0.36 & -1.83 & 15.23 \\
152 & -0.29 & -4.46 & 0.33 & -1.76 & 15.12 \\
153 & -0.33 & -4.34 & 0.36 & -1.80 & 15.24 \\
154 & -0.35 & -4.41 & 0.41 & -1.82 & 15.17 \\
155 & -0.32 & -4.34 & 0.31 & -1.79 & 15.24 \\
156 & -0.33 & -4.54 & 0.30 & -1.80 & 15.04 \\
157 & -0.28 & -4.64 & 0.36 & -1.75 & 14.94 \\
158 & -0.32 & -4.62 & 0.37 & -1.79 & 14.96 \\
159 & -0.33 & -4.64 & 0.35 & -1.8 & 14.94 \\
160 & -0.31 & -4.40 & 0.41 & -1.78 & 15.18 \\
161 & -0.32 & -4.61 & 0.33 & -1.79 & 14.97 \\
162 & -0.30 & -4.64 & 0.39 & -1.77 & 14.94 \\
163 & -0.34 & -4.47 & 0.34 & -1.81 & 15.11 \\
164 & -0.24 & -4.98 & 0.42 & -1.71 & 14.6 \\
165 & -0.29 & -4.70 & 0.43 & -1.76 & 14.88 \\
166 & -0.14 & -5.99 & 0.40 & -1.61 & 13.59 \\
167 & -0.26 & -4.98 & 0.31 & -1.73 & 14.60 \\
168 & -0.15 & -5.54 & 0.48 & -1.62 & 14.04 \\
169 & -0.25 & -4.76 & 0.53 & -1.72 & 14.82 \\
170 & -0.30 & -4.49 & 0.30 & -1.77 & 15.09 \\
171 & -0.36 & -4.18 & 0.30 & -1.83 & 15.40 \\
172 & -0.37 & -4.36 & 0.37 & -1.84 & 15.22 \\
173 & -0.31 & -4.47 & 0.28 & -1.78 & 15.11 \\
174 & -0.35 & -4.55 & 0.35 & -1.82 & 15.03 \\
175 & -0.31 & -4.49 & 0.43 & -1.78 & 15.09 \\
176 & -0.35 & -4.36 & 0.32 & -1.82 & 15.22 \\
177 & -0.35 & -4.26 & 0.45 & -1.82 & 15.32 \\
178 & -0.33 & -4.61 & 0.39 & -1.80 & 14.97 \\
179 & -0.27 & -4.78 & 0.43 & -1.74 & 14.80 \\
180 & -0.32 & -4.73 & 0.36 & -1.79 & 14.85 \\
181 & -0.29 & -4.69 & 0.3 & -1.76 & 14.89 \\
182 & -0.34 & -4.51 & 0.33 & -1.81 & 15.07 \\
183 & -0.28 & -4.93 & 0.36 & -1.75 & 14.65 \\
184 & -0.20 & -5.09 & 0.63 & -1.67 & 14.49 \\
185 & -0.30 & -4.90 & 0.34 & -1.77 & 14.68 \\
186 & -0.17 & -5.72 & 0.48 & -1.64 & 13.86 \\
187 & -0.28 & -4.78 & 0.38 & -1.75 & 14.80 \\
188 & -0.17 & -5.95 & 0.66 & -1.64 & 13.63 \\
189 & -0.29 & -4.86 & 0.36 & -1.76 & 14.72 \\
190 & -0.21 & -5.42 & 0.47 & -1.68 & 14.16 \\
191 & -0.33 & -4.47 & 0.26 & -1.80 & 15.11 \\
192 & -0.35 & -4.67 & 0.39 & -1.82 & 14.91 \\
193 & -0.35 & -4.30 & 0.28 & -1.82 & 15.28 \\
194 & -0.37 & -4.18 & 0.31 & -1.84 & 15.40 \\
195 & -0.37 & -4.51 & 0.37 & -1.84 & 15.07 \\
196 & -0.34 & -4.31 & 0.34 & -1.81 & 15.27 \\
197 & -0.38 & -4.35 & 0.32 & -1.85 & 15.23 \\
198 & -0.35 & -4.22 & 0.34 & -1.82 & 15.36 \\
199 & -0.32 & -4.74 & 0.35 & -1.79 & 14.84 \\
200 & -0.32 & -4.64 & 0.38 & -1.79 & 14.94 \\
201 & -0.32 & -4.89 & 0.3 & -1.79 & 14.69 \\
202 & -0.33 & -4.56 & 0.29 & -1.80 & 15.02 \\
203 & -0.35 & -4.55 & 0.32 & -1.82 & 15.03 \\
204 & -0.29 & -4.73 & 0.41 & -1.76 & 14.85 \\
205 & -0.26 & -4.90 & 0.34 & -1.73 & 14.68 \\
206 & -0.31 & -4.67 & 0.37 & -1.78 & 14.91 \\
207 & -0.27 & -6.46 & 0.43 & -1.74 & 13.12 \\
208 & -0.17 & -5.93 & 0.52 & -1.64 & 13.65 \\
209 & -0.21 & -5.60 & 0.57 & -1.68 & 13.98 \\
210 & -0.28 & -6.25 & 0.31 & -1.75 & 13.33 \\
211 & -0.26 & -4.93 & 0.45 & -1.73 & 14.65 \\
213 & -0.37 & -4.12 & 0.31 & -1.84 & 15.46 \\
215 & -0.40 & -4.42 & 0.31 & -1.87 & 15.16 \\
216 & -0.34 & -4.44 & 0.35 & -1.81 & 15.14 \\
217 & -0.31 & -4.64 & 0.31 & -1.78 & 14.94 \\
218 & -0.35 & -4.70 & 0.32 & -1.82 & 14.88 \\
219 & -0.34 & -4.38 & 0.27 & -1.81 & 15.20 \\
220 & -0.37 & -4.47 & 0.34 & -1.84 & 15.11 \\
221 & -0.38 & -4.36 & 0.28 & -1.85 & 15.22 \\
222 & -0.36 & -4.36 & 0.33 & -1.83 & 15.22 \\
223 & -0.30 & -4.69 & 0.36 & -1.77 & 14.89 \\
224 & -0.38 & -4.37 & 0.32 & -1.85 & 15.21 \\
225 & -0.27 & -4.87 & 0.45 & -1.74 & 14.71 \\
226 & -0.28 & -4.91 & 0.35 & -1.75 & 14.67 \\
227 & -0.30 & -4.93 & 0.31 & -1.77 & 14.65 \\
229 & -0.32 & -4.57 & 0.38 & -1.79 & 15.01 \\
230 & -0.30 & -4.93 & 0.36 & -1.77 & 14.65 \\
231 & -0.31 & -6.31 & 0.26 & -1.78 & 13.27 \\
234 & -0.25 & -4.94 & 0.44 & -1.72 & 14.64 \\
236 & -0.34 & -4.46 & 0.36 & -1.81 & 15.12 \\
237 & -0.39 & -4.36 & 0.25 & -1.86 & 15.22 \\
238 & -0.39 & -4.33 & 0.27 & -1.86 & 15.25 \\
239 & -0.36 & -4.31 & 0.34 & -1.83 & 15.27 \\
240 & -0.35 & -4.56 & 0.28 & -1.82 & 15.02 \\
241 & -0.38 & -4.38 & 0.30 & -1.85 & 15.20 \\
242 & -0.35 & -4.37 & 0.37 & -1.82 & 15.21 \\
243 & -0.37 & -4.44 & 0.28 & -1.84 & 15.14 \\
244 & -0.37 & -4.44 & 0.35 & -1.84 & 15.14 \\
245 & -0.36 & -4.45 & 0.31 & -1.83 & 15.13 \\
246 & -0.32 & -4.71 & 0.28 & -1.79 & 14.87 \\
247 & -0.33 & -4.69 & 0.34 & -1.80 & 14.89 \\
250 & -0.37 & -4.46 & 0.28 & -1.84 & 15.12 \\
251 & -0.30 & -4.75 & 0.29 & -1.77 & 14.83 \\
252 & -0.32 & -4.61 & 0.41 & -1.79 & 14.97 \\
253 & -0.28 & -4.95 & 0.47 & -1.75 & 14.63 \\
255 & -0.32 & -4.70 & 0.35 & -1.79 & 14.88 \\
256 & -0.25 & -5.07 & 0.35 & -1.72 & 14.51 \\
258 & -0.40 & -4.26 & 0.36 & -1.87 & 15.32 \\
259 & -0.41 & -4.28 & 0.35 & -1.88 & 15.30 \\
260 & -0.41 & -4.29 & 0.29 & -1.88 & 15.29 \\
261 & -0.37 & -4.33 & 0.35 & -1.84 & 15.25 \\
262 & -0.36 & -4.47 & 0.29 & -1.83 & 15.11 \\
263 & -0.37 & -4.56 & 0.31 & -1.84 & 15.02 \\
264 & -0.35 & -4.53 & 0.29 & -1.82 & 15.05 \\
267 & -0.38 & -4.45 & 0.28 & -1.85 & 15.13 \\
268 & -0.40 & -4.26 & 0.33 & -1.87 & 15.32 \\
269 & -0.36 & -4.49 & 0.28 & -1.83 & 15.09 \\
270 & -0.37 & -4.45 & 0.30 & -1.84 & 15.13 \\
271 & -0.37 & -4.43 & 0.27 & -1.84 & 15.15 \\
272 & -0.36 & -4.53 & 0.34 & -1.83 & 15.05 \\
273 & -0.36 & -4.52 & 0.31 & -1.83 & 15.06 \\
274 & -0.27 & -5.27 & 0.42 & -1.74 & 14.31 \\
275 & -0.31 & -4.94 & 0.49 & -1.78 & 14.64 \\
276 & -0.27 & -5.04 & 0.44 & -1.74 & 14.54 \\
277 & -0.28 & -4.94 & 0.46 & -1.75 & 14.64 \\
278 & -0.35 & -4.63 & 0.28 & -1.82 & 14.95 \\
280 & -0.40 & -4.30 & 0.29 & -1.87 & 15.28 \\
281 & -0.42 & -4.25 & 0.32 & -1.89 & 15.33 \\
282 & -0.35 & -4.58 & 0.32 & -1.82 & 15.00 \\
283 & -0.37 & -4.44 & 0.39 & -1.84 & 15.14 \\
284 & -0.40 & -4.50 & 0.29 & -1.87 & 15.08 \\
286 & -0.33 & -4.97 & 0.31 & -1.80 & 14.61 \\
287 & -0.36 & -4.46 & 0.30 & -1.83 & 15.12 \\
288 & -0.41 & -4.39 & 0.29 & -1.88 & 15.19 \\
289 & -0.39 & -4.47 & 0.31 & -1.86 & 15.11 \\
290 & -0.40 & -4.28 & 0.30 & -1.87 & 15.30 \\
291 & -0.39 & -4.41 & 0.27 & -1.86 & 15.17 \\
292 & -0.39 & -4.34 & 0.32 & -1.86 & 15.24 \\
293 & -0.34 & -4.65 & 0.39 & -1.81 & 14.93 \\
294 & -0.39 & -4.49 & 0.33 & -1.86 & 15.09 \\
295 & -0.38 & -4.55 & 0.30 & -1.85 & 15.03 \\
296 & -0.34 & -5.02 & 0.42 & -1.81 & 14.56 \\
297 & -0.31 & -4.93 & 0.34 & -1.78 & 14.65 \\
298 & -0.30 & -4.88 & 0.32 & -1.77 & 14.70 \\
299 & -0.27 & -4.95 & 0.38 & -1.74 & 14.63 \\
300 & -0.40 & -4.23 & 0.32 & -1.87 & 15.35 \\
301 & -0.34 & -5.56 & 0.29 & -1.81 & 14.02 \\
302 & -0.43 & -4.36 & 0.27 & -1.9 & 15.22 \\
303 & -0.38 & -4.49 & 0.25 & -1.85 & 15.09 \\
304 & -0.39 & -4.40 & 0.31 & -1.86 & 15.18 \\
305 & -0.43 & -4.42 & 0.32 & -1.90 & 15.16 \\
306 & -0.30 & -5.63 & 0.28 & -1.77 & 13.95 \\
307 & -0.35 & -4.84 & 0.34 & -1.82 & 14.74 \\
308 & -0.41 & -4.34 & 0.40 & -1.88 & 15.24 \\
309 & -0.39 & -4.60 & 0.32 & -1.86 & 14.98 \\
310 & -0.41 & -4.47 & 0.27 & -1.88 & 15.11 \\
311 & -0.39 & -4.47 & 0.26 & -1.86 & 15.11 \\
312 & -0.38 & -4.67 & 0.32 & -1.85 & 14.91 \\
313 & -0.40 & -4.34 & 0.30 & -1.87 & 15.24 \\
314 & -0.35 & -4.63 & 0.29 & -1.82 & 14.95 \\
315 & -0.37 & -4.52 & 0.29 & -1.84 & 15.06 \\
316 & -0.38 & -4.53 & 0.23 & -1.85 & 15.05 \\
317 & -0.36 & -5.99 & 0.20 & -1.83 & 13.59 \\
318 & -0.34 & -4.73 & 0.35 & -1.81 & 14.85 \\
319 & -0.35 & -4.78 & 0.27 & -1.82 & 14.8 \\
320 & -0.36 & -4.63 & 0.30 & -1.83 & 14.95 \\
321 & -0.40 & -4.47 & 0.34 & -1.87 & 15.11 \\
322 & -0.34 & -5.27 & 0.34 & -1.81 & 14.31 \\
323 & -0.42 & -4.46 & 0.29 & -1.89 & 15.12 \\
324 & -0.42 & -4.29 & 0.31 & -1.89 & 15.29 \\
325 & -0.39 & -4.53 & 0.27 & -1.86 & 15.05 \\
326 & -0.42 & -4.49 & 0.26 & -1.89 & 15.09 \\
327 & -0.35 & -4.74 & 0.30 & -1.82 & 14.84 \\
328 & -0.32 & -4.90 & 0.31 & -1.79 & 14.68 \\
329 & -0.40 & -4.45 & 0.29 & -1.87 & 15.13 \\
330 & -0.44 & -4.32 & 0.36 & -1.91 & 15.26 \\
331 & -0.45 & -4.15 & 0.28 & -1.92 & 15.43 \\
332 & -0.43 & -4.31 & 0.31 & -1.90& 15.27 \\
333 & -0.43 & -4.38 & 0.29 & -1.90 & 15.20 \\
334 & -0.45 & -4.20 & 0.35 & -1.92 & 15.38 \\
335 & -0.41 & -4.49 & 0.29 & -1.88 & 15.09 \\
336 & -0.39 & -4.41 & 0.29 & -1.86 & 15.17 \\
337 & -0.42 & -4.31 & 0.29 & -1.89 & 15.27 \\
338 & -0.38 & -4.60 & 0.25 & -1.85 & 14.98 \\
339 & -0.37 & -4.65 & 0.29 & -1.84 & 14.93 \\
340 & -0.36 & -4.73 & 0.28 & -1.83 & 14.85 \\
341 & -0.42 & -4.37 & 0.31 & -1.89 & 15.21 \\
342 & -0.42 & -4.31 & 0.30 & -1.89 & 15.27 \\
343 & -0.37 & -4.71 & 0.34 & -1.84 & 14.87 \\
344 & -0.40 & -4.57 & 0.33 & -1.87 & 15.01 \\
345 & -0.45 & -4.35 & 0.27 & -1.92 & 15.23 \\
346 & -0.38 & -4.58 & 0.34 & -1.85 & 15.00 \\
347 & -0.35 & -4.70 & 0.33 & -1.82 & 14.88 \\
348 & -0.45 & -4.30 & 0.22 & -1.92 & 15.28 \\
349 & -0.39 & -4.69 & 0.27 & -1.86 & 14.89 \\
350 & -0.45 & -4.33 & 0.27 & -1.92 & 15.25 \\
351 & -0.45 & -4.26 & 0.28 & -1.92 & 15.32 \\
352 & -0.40 & -4.63 & 0.23 & -1.87 & 14.95 \\
353 & -0.41 & -4.48 & 0.29 & -1.88 & 15.10 \\
354 & -0.41 & -4.70 & 0.26 & -1.88 & 14.88 \\
355 & -0.41 & -4.51 & 0.32 & -1.88 & 15.07 \\
356 & -0.44 & -4.27 & 0.20 & -1.91 & 15.31 \\
357 & -0.39 & -4.60 & 0.26 & -1.86 & 14.98 \\
358 & -0.40 & -4.50 & 0.30 & -1.87 & 15.08 \\
359 & -0.41 & -4.48 & 0.32 & -1.88 & 15.10 \\
360 & -0.43 & -4.41 & 0.28 & -1.90 & 15.17 \\
361 & -0.41 & -4.45 & 0.29 & -1.88 & 15.13 \\
362 & -0.42 & -4.51 & 0.31 & -1.89 & 15.07 \\
363 & -0.48 & -4.26 & 0.22 & -1.95 & 15.32 \\
364 & -0.40 & -4.58 & 0.25 & -1.87 & 15.00 \\
365 & -0.37 & -4.72 & 0.32 & -1.84 & 14.86 \\
366 & -0.46 & -4.32 & 0.32 & -1.93 & 15.26 \\
367 & -0.46 & -4.36 & 0.29 & -1.93 & 15.22 \\
368 & -0.44 & -4.54 & 0.24 & -1.91 & 15.04 \\
369 & -0.45 & -4.39 & 0.31 & -1.92 & 15.19 \\
370 & -0.44 & -4.54 & 0.28 & -1.91 & 15.04 \\
371 & -0.43 & -4.48 & 0.28 & -1.9 & 15.10 \\
372 & -0.46 & -4.44 & 0.28 & -1.93 & 15.14 \\
373 & -0.42 & -4.56 & 0.22 & -1.89 & 15.02 \\
374 & -0.44 & -4.36 & 0.25 & -1.91 & 15.22 \\
375 & -0.42 & -4.55 & 0.31 & -1.89 & 15.03 \\
376 & -0.46 & -4.30 & 0.25 & -1.93 & 15.28 \\
377 & -0.46 & -4.35 & 0.29 & -1.93 & 15.23 \\
378 & -0.45 & -4.53 & 0.31 & -1.92 & 15.05 \\
379 & -0.43 & -4.47 & 0.28 & -1.90 & 15.11 \\
380 & -0.46 & -4.28 & 0.32 & -1.93 & 15.30 \\
381 & -0.44 & -4.47 & 0.30 & -1.91 & 15.11 \\
382 & -0.40 & -4.81 & 0.36 & -1.87 & 14.77 \\
383 & -0.43 & -4.49 & 0.27 & -1.90 & 15.09 \\
384 & -0.45 & -4.53 & 0.36 & -1.92 & 15.05 \\
385 & -0.43 & -4.66 & 0.22 & -1.9 & 14.92 \\
386 & -0.45 & -4.47 & 0.29 & -1.92 & 15.11 \\
387 & -0.45 & -4.51 & 0.26 & -1.92 & 15.07 \\
388 & -0.43 & -4.48 & 0.30 & -1.90 & 15.10 \\
389 & -0.45 & -4.52 & 0.25 & -1.92 & 15.06 \\
390 & -0.48 & -4.38 & 0.28 & -1.95 & 15.2 \\
391 & -0.44 & -4.45 & 0.29 & -1.91 & 15.13 \\
392 & -0.44 & -4.52 & 0.32 & -1.91 & 15.06 \\
393 & -0.43 & -4.56 & 0.23 & -1.90 & 15.02 \\
394 & -0.43 & -4.63 & 0.27 & -1.90 & 14.95 \\
395 & -0.42 & -5.00 & 0.60 & -1.89 & 14.58 \\
396 & -0.43 & -4.61 & 0.30 & -1.90 & 14.97 \\
397 & -0.46 & -4.41 & 0.33 & -1.93 & 15.17 \\
398 & -0.47 & -4.94 & 0.74 & -1.94 & 14.64 \\
399 & -0.45 & -4.56 & 0.32 & -1.92 & 15.02 \\
400 & -0.47 & -4.52 & 0.28 & -1.94 & 15.06 \\
401 & -0.47 & -4.47 & 0.28 & -1.94 & 15.11 \\
402 & -0.44 & -4.81 & 0.43 & -1.91 & 14.77 \\
403 & -0.47 & -4.39 & 0.27 & -1.94 & 15.19 \\
404 & -0.42 & -4.8 & 0.28 & -1.89 & 14.78 \\
405 & -0.45 & -4.66 & 0.28 & -1.92 & 14.92 \\
406 & -0.46 & -4.54 & 0.28 & -1.93 & 15.04 \\
  \caption{Model parameters for each field of the survey.}
        \end{longtable}
\endgroup

\end{appendix}

\end{document}